\documentclass[chicago]{emulateapj_jw}

\usepackage{hyperref}
\usepackage{epsfig}
\usepackage{natbib}
\usepackage{graphicx}                
\usepackage{lscape}
\usepackage{verbatim}              
\usepackage{amsmath,amsthm,amsfonts,amsopn,amssymb} 
\usepackage{longtable}
\usepackage{pdflscape}
\usepackage{afterpage}
\usepackage{rotating}					
\usepackage{ulem}
\usepackage{epstopdf}  
\usepackage{multirow}
\usepackage{color}
\usepackage[left]{lineno}
\newcommand{\totaltargets}{84 }
\newcommand{\totalplanets}{97 }
\newcommand{\totalmulti}{27 }
\newcommand{\detectstar}{59 }
\newcommand{\detectsys}{40 }
\newcommand{\rvstar}{19 }
\newcommand{\myaostar}{60 }
\newcommand{\myaopalomar}{40 }
\newcommand{\myaokeck}{27 }
\newcommand{\myaonewstar}{29 }
\newcommand{\myaonewsys}{22 }
\newcommand{\myaonewcolor}{8 }
\newcommand{\myaomiss}{11 }
\newcommand{\multicolor}{21 }
\newcommand{\multicolorod}{6 }
\newcommand{\multicolorsubarc}{5 }
\newcommand{\singleall}{38 }
\newcommand{\singleK}{29 }
\newcommand{\starK}{51 }

\received{}
\accepted{}

\slugcomment{to appear in ApJ}
\shorttitle{}
\shortauthors{Wang et al.}

\begin{document}
\title{Influence of Stellar Multiplicity On Planet Formation. III. Adaptive Optics Imaging of Kepler Stars With Gas Giant Planets}
\author{
Ji Wang\altaffilmark{1},
Debra A. Fischer\altaffilmark{1},
Elliott P. Horch\altaffilmark{2} and
Ji-Wei Xie\altaffilmark{3},
} 
\email{ji.wang@yale.edu}
\altaffiltext{1}{Department of Astronomy, Yale University, New Haven, CT 06511 USA}
\altaffiltext{2}{Department of Physics, Southern Connecticut State University, 501 Crescent Street, New Haven, CT 06515, USA}
\altaffiltext{3}{Department of Astronomy \& Key Laboratory of Modern Astronomy and Astrophysics in Ministry of Education, Nanjing University,
210093, China}

\begin{abstract}

As hundreds of gas giant planets have been discovered, we study how these planets form and evolve in different stellar environments, specifically in multiple stellar systems. In such systems, stellar companions may have a profound influence on gas giant planet formation and evolution via several dynamical effects such as truncation and perturbation. We select \totaltargets Kepler Objects of Interest (KOIs) with gas giant planet candidates. We obtain high-angular resolution images using telescopes with adaptive optics (AO) systems. Together with the AO data, we use archival radial velocity data and dynamical analysis to constrain the presence of stellar companions. We detect \detectstar stellar companions around \detectsys KOIs for which we develop methods of testing their physical association. These methods are based on color information and galactic stellar population statistics. We find evidence of suppressive planet formation within 20 AU by comparing stellar multiplicity. The stellar multiplicity rate for planet host stars is 0$^{+5}_{-0}$\% within 20 AU. In comparison, the stellar multiplicity rate is 18\%$\pm$2\% for the control sample, i.e., field stars in the solar neighborhood. The stellar multiplicity rate for planet host stars is 34\%$\pm$8\% for separations between 20 and 200 AU, which is higher than the control sample at 12\%$\pm$2\%. Beyond 200 AU, stellar multiplicity rates are comparable between planet host stars and the control sample. We discuss the implications of the results to gas giant planet formation and evolution.   

\end{abstract}


\section{Introduction}
\label{sec:intro}
Almost half of sun-like stars are members of binary or multiple star systems~\citep[hereafter MSS,][]{Raghavan2010,Duquennoy1991}. During the formation of a star-planet system, the presence of a companion star is likely to perturb or truncate the protoplanetary disk~\citep[e.g.,][]{HW99,JangCondell2007}. These effects would affect the formation rate of planets around the stellar components of MSS. Indeed, observations of star forming regions show that protoplanetary disks around MSS have shorter lifetimes than disks around single star systems~\citep[hereafter SSS,][]{Kraus2012}. Furthermore, after planet formation, dynamical interactions with stellar companions could affect the orbital evolution of a planet and its survival rate via perturbation~\citep{Wu2003,Naoz2012}, disk-driven migration~\citep{Lin1986,Tanaka2002} and planet-planet scattering~\citep{Rasio1996,Chatterjee2008}.

Despite many theoretical and numerical studies of planets in MSS, the planet occurrence rate in MSS is still uncertain. There have been numerous estimates of the planet occurrence rate for solar-type stars based on the Kepler results~\citep[e.g.,][]{Howard2012,Fressin2013}; however, these studies do not distinguish between planet host stars in SSS or MSS. The lack of stellar multiplicity information prevents us from comparing the planet occurrence rate between SSS and MSS. The comparison would provide insights into the influence of stellar companions on planet formation. The discovery of thousands of planet candidates from the Kepler mission~\citep{Borucki2011,Batalha2013,Burke2014} provides a unique opportunity to study planets in MSS~\citep[e.g.,][]{Wang2014b}

There are two ways of studying planets in MSS. First, one can select a sample of known MSS and then search for planets around each of the component stars~\citep[the Planet-Quest approach, e.g.,][]{Konacki2005, Eggenberger2007, Konacki2009,Toyota2009}. Unfortunately, for ground based surveys, the detection efficiency of planets in MSS is affected by flux contamination from the additional stellar components~\citep{Wright2012}. Flux contamination also reduces the signal to noise ratio (SNR) of a transiting planet~\citep{Fressin2013,Wang2014b}. On the other hand, the Star-Quest approach is an effective way to study planets in MSS. Using a sample of known stars with planets, it is possible to determine the fraction of MSS in that sample~\citep[e.g.,][]{Luhman2002,Patience2002,Wang2014a,Ngo2014}. From a technical standpoint, it is much easier to detect stellar companions than planetary companions, making the Star-Quest approach much easier and more sensitive to lower mass planets. 
Previous observational studies have suggested that the presence of a stellar companion suppresses planet formation~\citep[e.g.,][]{Eggenberger2011}. However, these studies have not fully considered biases in target selection and planet detection or the observational incompleteness in the search for stellar companions of planet host stars. ~\citet{Wang2014b} discuss the above concerns and find that planet formation is suppressed by stellar companions with separations up to 1500 AU.

~\citet{Wang2014a,Wang2014b} summarize the works following the Star-Quest approach prior to 2014. Since then, more progress has been made. These new results suggest that visual stellar companions are not rare around planet host stars. Using the Lucky Imaging technique, ~\citet{LilloBox2014} find that 32.8\% of Kepler planet host stars have at least one visual stellar companion within 6$^{\prime\prime}$. Based on adaptive optics (AO) imaging of 87 Kepler planet host stars,~\citet{Dressing2014} find that 31.0\% of planet host stars have stellar companions within 4$^{\prime\prime}$.~\citet{Law2014} observe 715 Kepler stars with planet candidates with the Robo-AO system. They find 7.4\% of Kepler planet host stars have stellar companions within 2.5$^{\prime\prime}$. However, the fraction of gravitationally bound companions is uncertain from these surveys. Using the speckle imaging technique, ~\citet{Horch2014} detect 49 stellar companions within 1 arcsec around over 600 Kepler stars with planet candidates. The majority of the detected companions are likely to be gravitationally bound to the planet host stars based on a statistical argument. Accounting for detection incompleteness, they conclude that the stellar multiplicity rate for planet host stars is similar to stars in the solar neighborhood.~\citet{Gilliland2014} observe 23 Kepler stars with small and cool planet candidates using the Hubble Space Telescope. They find evidence that physically-associated stellar companions are more common around planet host stars than around field stars in the solar neighborhood. The Kepler stars in the aforementioned studies have an average Kepler magnitude of $13.5$. Assuming that they are solar-type stars, the average distance is $\sim$600-700 pc. More recently,~\citet{Ngo2014} focus on a sample of host stars for hot Jupiters (HJs) that were discovered and confirmed by ground-based observations. They use AO imaging and conduct multiple-epoch observations to confirm physical association by measuring common proper motions of stellar components. The stellar multiplicity rate for HJ host stars is almost twice as high as that for stars in the solar neighborhood for stellar separations between 50 and 2000 AU. The overabundance of stellar companions for HJ host stars suggests the positive role of stellar companions in HJ formation and evolution. 

While rapid progress has been made since 2014, there are still a number of issues for the Star-Quest approach. First, there is a selection bias against MSS for work on planets detected by ground-based observations. MSS with small separations and considerable flux contamination are usually excluded in ground-based surveys for planets~\citep[e.g.,][]{Wright2012}. The selection bias is difficult to quantify and correct, so converting the measured stellar multiplicity rate to planet occurrence rate is challenging~\citep{Wang2014b}. Second, most studies focus on a single detection technique for stellar companions. The high-angular resolution imaging technique has been the dominant method. However, the physical association of stellar companions is difficult to assess for Kepler stars. This becomes an issue when calculating the stellar multiplicity rate which concerns only gravitationally bound companions. Third, the imaging techniques are not effective in detecting stellar companions within or near the diffraction limits of telescopes, which correspond to 20-50 AU in physical separation. Stellar companions at smaller separations can be more effectively detected by measuring the radial velocity (RV). So far, only a few studies following the Star-Quest approach use the RV technique in combination with the high-angular resolution imaging technique, which dramatically increases the search completeness for stellar companions~\citep{Knutson2013,Wang2014b,Ngo2014}.  

Finally, the control sample to be compared is not a perfect sample. Field stars in the solar neighborhood usually serve as a control sample for stars without planets, but studies on planet occurrence rate suggest that the majority of stars host at least one planet~\citep[e.g.,][]{Mayor2011,Howard2012,Fressin2013,Dressing2013}. In order to construct a more meaningful control sample, we can find other stars that do not have planets down to a certain mass and up to a certain orbital separation~\citep[e.g.,][]{Eggenberger2007}, or we can continue to use the field stars as a control sample but set a planet mass/radius and separation range to limit the level of contamination of planet host stars. For example, if we consider only gas giant planets within $\sim$1 AU, then the stars in the solar neighborhood can serve as a reasonable control sample because fewer than 10\% of stars have gas giant planets within 1 AU~\citep{Cumming2008,Mayor2011}.

To address the above issues, we have conducted a search for stellar companions for \totaltargets Kepler stars with gas giant planets within $\sim$1 AU. This sample of stars is not biased against MSS because the Kepler mission does not apply a selection criterion that excludes MSS~\citep{Brown2011}. The typical full width at half maximum (FWHM) of images for Kepler target selection is 2.5$^{\prime\prime}$ and thus these images are ineffective at distinguishing binary stars. We use both RV data and high-angular resolution imaging data for these stars, so the search completeness is high compared to surveys employing only one technique. We assess the physical association of detected stellar companions using (1) their color information and (2) a statistical argument based on a galactic stellar population model. We can compare the stellar multiplicity rate for this sample to that for field stars in the solar neighborhood without a significant contamination of planet host stars in the control sample. After these issues are resolved, we can study the influence of stellar companions on gas giant planet formation with the sample \totaltargets Kepler planet host stars. 

The paper is organized as follows. We describe the sample of Kepler stars with gas giant planets in \S \ref{sec:Sample}. AO observation and data reduction for these stars are presented in \S \ref{sec:Search}. We also discuss the physical association of detected stellar companions in \S \ref{sec:phy_ass}. We synthesize the results of different techniques in the search for stellar companions in \S \ref{sec:aorvda}. These techniques include the RV and AO imaging techniques and the dynamical analysis. The stellar multiplicity rate of Kepler stars with gas giant planets is given in \S \ref{sec:planet_frequency}. Discussion and summary are given in \S \ref{sec:discussion} and \S \ref{sec:Summary}.

\section{Sample Description}
\label{sec:Sample}

From the NASA Exoplanet Archive\footnote{http://exoplanetarchive.ipac.caltech.edu}, we select Kepler Objects of Interest (KOIs) that satisfy the following criteria: (1), disposition of either Candidate or Confirmed; (2), stellar effective temperature (T$\rm{eff}$) lower than 6500 K; (3) stellar surface gravity ($\log g$) higher than 4.0; (4), Kepler magnitude ($K_P$) brighter than 14th mag; (5), with at least one gas giant planet ($3.8\ R_\oplus\le R_P\le 22.0\ R_\oplus$). In total, we select \totaltargets KOIs with \totalplanets gas giant planets. Stellar and orbital parameters for these KOIs are given in Table \ref{tab:stellar_params}. The median distance of these KOIs is 580 pc. There are \totalmulti multi-planet systems among \totaltargets KOIs. 

There are \rvstar KOIs with RV observations. We obtain RV data from the Kepler Community Follow-up Observation Program\footnote{https://cfop.ipac.caltech.edu} (CFOP). The majority of the RV data (14 out of \rvstar) were taken with the HIRES instrument~\citep{Vogt1994} and reported in~\citet{Marcy2014}. Exceptions are KOI-1~\citep[TrES-2,][]{Donovan2006}, KOI-3~\citep[HAT-P-11 b,][]{Bakos2010}, KOI-97~\citep[Kepler-7 b,][]{Latham2010}, KOI-128~\citep[Kepler-15 b,][]{Endl2011}, and KOI-135~\citep[Kepler-43 b,][]{Bonomo2012}. The Modified Julian Dates (MJDs) of the first and last RV data points and the number of RV data points for each KOI are given in Table \ref{tab:stellar_params}. 

For KOIs with high-angular resolution images from CFOP, we use the images to search for stellar companions. These images were taken at different telescopes including Keck, Palomar, MMT, Lick, and WIYN. For KOIs whose high-angular resolution images are not available, we have taken AO images using the PHARO (Palomar High Angular Resolution Observer) instrument~\citep{Brandl1997,Hayward2001} at the Palomar 200-inch telescope and the NIRC2 instrument~\citep{Wizinowich2000} at the Keck II telescope. In total, we have taken AO images for \myaostar out of the \totaltargets KOIs. Telescope and photometric band information is given in Table \ref{tab:stellar_params}. The KOIs with AO images taken through this work are also indicated in Table \ref{tab:stellar_params}.

\section{AO Observation and Data Reduction}
\label{sec:Search}

\subsection{AO Imaging with PHARO at Palomar}
\label{sec:pharo}

We observed \myaopalomar KOIs in the sample with the PHARO instrument\citep{Brandl1997,Hayward2001} at the Palomar 200-inch telescope. The observations were made between UT July 13rd and 17th in 2014 with seeing varying between 1.0$^{\prime\prime}$ and 2.5$^{\prime\prime}$. PHARO is behind the Palomar-3000 AO system, which provides a on-sky Strehl of 86\% in $K$ band~\citep{Burruss2014}. The pixel scale of PHARO is 25 mas pixel$^{-1}$. With a mosaic 1K $\times$1K detector, the field of view (FOV) is 25$^{\prime\prime}\times$25$^{\prime\prime}$. We normally obtained the first image in $K$ band with a 5-point dither pattern, which had a throw of 2.5$^{\prime\prime}$. The exposure time was set such that the peak flux of the KOI is at least 10,000 ADU for each frame, which is within the linear range of the detector. If a stellar companion was detected, we observed the KOI in $J$ and $H$ bands. 

\subsection{AO Imaging with NIRC2 at Keck II}
\label{sec:nirc2}

We observed \myaokeck KOIs in the sample with the NIRC2 instrument~\citep{Wizinowich2000} at the Keck II telescope. The observations were made on UT July 18th and August 18th in 2014 with excellent/good seeing between 0.3$^{\prime\prime}$ to 0.8$^{\prime\prime}$. NIRC2 is a near infrared imager designed for the Keck AO system. We selected the narrow camera mode, which has a pixel scale of 10 mas pixel$^{-1}$. The FOV is thus 10$^{\prime\prime}\times$10$^{\prime\prime}$ for a mosaic 1K $\times$1K detector. We started the observation in $K$ band for each KOI. The exposure time setting is the same as the PHARO observation: we ensured that the peak flux is at least 10,000 ADU for each frame. We used a 3-point dither pattern with a throw of 2.5$^{\prime\prime}$. We avoided the lower left quadrant in the dither pattern because it has a much higher instrumental noise than other 3 quadrants on the detector. We continued observations of a KOI in $J$ and $H$ bands if any stellar companions were found. 

\subsection{Contrast Curve and Detections}
\label{sec:contrast_curve}

The raw data were processed using standard techniques to replace bad pixels, flat-field, subtract thermal background, align and co-add frames. We calculated the 5-$\sigma$ detection limit as follows. We defined a series of concentric annuli centering on the star. For the concentric annuli, we calculated the median and the standard deviation of flux for pixels within these annuli. We used the value of five times the standard deviation above the median as the 5-$\sigma$ detection limit. The median contrast curve and the 1-$\sigma$ deviation of $K$ band AO images we used in this paper are shown in Fig. \ref{fig:AO_contrast}. Also plotted are detected stellar companions as indicated by asterisks in Fig. \ref{fig:AO_contrast}. These companions are brighter than the contrast curve, so the significance of detections is at least 5 $\sigma$. In total, \detectstar stellar companions were detected around \detectsys KOIs. Their stellar and orbital properties are summarized in Table \ref{tab:AO_params}.

\subsection{Comparison to Previous Work}
\label{sec:comp_prev}

Among \detectstar stellar companions, \myaonewstar are newly detected around \myaonewsys KOIs in this study. Furthermore, we add observations to \myaonewcolor previously known stellar companions in additional color filters. The other stellar companions that were previously reported are noted with references in Table \ref{tab:AO_params}. These observation campaigns were carried out using a variety of instruments at different telescopes, e.g., AIRES at MMT~\citep{Adams2012,Dressing2014}, PHARO at Palomar~\citep{Adams2012}, Robo-AO at Palomar~\citep{Law2014}, DSSI at WIYN and Gemini~\citep{Horch2014}, and AstraLux at Calar Alto~\citep{LilloBox2014}. When compared to previous work, we miss \myaomiss stellar companions. They are marked with an asterisk in Table \ref{tab:AO_params}. All but one (KOI-372, $\Delta K=4.0$) stellar companions that we miss are very faint, with a differential magnitude range between 7.2 and 8.2. Our pipeline does not identify these companions possibly because of different detection criteria. In one case (KOI-377, $\Delta J=6.8$), the stellar companion is identified in $K$ band, but not in $J$ band. 

\section{Physical Association}
\label{sec:phy_ass}

For stellar companions detected by imaging techniques, we need to confirm that they are not optical doubles/multiples. Otherwise, the unassociated stellar companions will systematically increase the stellar multiplicity rate and cause misinterpretations. To test physical association, the method of obtaining multiple-epoch images and measuring common proper motion has been proven effective~\citep{Ngo2014}. In our case, Kepler stars are generally $\sim$300-1000 pc away. While future observations are scheduled, common proper motion measurements are relatively more difficult. Given only one epoch of observation, we can use color information of detected stellar companions and assess the probability of their physical association to primary stars~\citep{LilloBox2014,Wang2014b}. Details of this approach are given in \S \ref{sec:phy_ass_color}. For stellar companions with only single-band observations, color information is not available. We can assess the probability with a galactic stellar population simulation (\S \ref{sec:phy_ass_galactic}). 

\subsection{Physical Association Based on Color Information}
\label{sec:phy_ass_color}

We compare the distance of a KOI and its stellar companions. If their distances do not match within uncertainty, then they are likely to be optical doubles and the physical association is excluded. For the distance of a KOI, we follow the method described in~\citet{Wang2014b}. We calculate the distant modulus for each KOI. The $V$ band apparent magnitude is obtained through the NASA Exoplanet Archive. The $V$ band absolute magnitude is calculated using the Yale-Yonsei (Y2) stellar evolution model~\citep{Demarque2004}. The input parameters for the Y2 model are T$_{\rm{eff}}$, $\log g$, age, and [Fe/H], which are also obtained through the NASA Exoplanet Archive. $V$ band extinction ($A_V$) is obtained from the Mikulski Archive for Space Telescopes\footnote{http://archive.stsci.edu/} (MAST). With the apparent and absolute $V$ band magnitudes and the extinction $A_V$, we can calculate the distance modulus of a KOI and thus its distance. For those KOIs whose extinctions are not available, we use $K$ band distance modulus assuming zero extinction in $K$ band. 

For stellar companions around a KOI, we use the color information, if available, to estimate their distances. We have color information, i.e., multi-band detections, for \multicolor companions. We convert the differential magnitudes to the true color of the companion. Based on the color information, we estimate the effective temperature of a stellar companion using Table 5 in~\citet{Kraus2007}. For stellar companions detected in more than 2 bands, we use the mean effective temperatures weighted by uncertainties. Once the effective temperature is available, we can find the corresponding $K$ band absolute magnitude for a stellar companion. Its apparent $K$ band magnitude can be calculated from the differential $K$ band magnitude and the apparent $K$ band magnitude of the KOI. The $K$ band distance modulus can be calculated assuming zero extinction. The distance modulus can then be used to estimate the distance of a stellar companion.

In the above calculation, extinctions in different bands need to be considered. Otherwise, a stellar companion would appear redder and closer. To account for extinction in different bands, we use a linear relation between $A_\lambda/A_V$ and 1/$\lambda$~\citep{Gordon2003}. Since we are only interested in the wavelength region between 0.55 $\mu m$ ($V$ band) and 2.19 $\mu m$ ($K$ band), a linear relation is a reasonable approximation. On one end, we assume $K$ band extinction to be zero. On the other end, we use the $A_V$ from the MAST archive. Extinctions in $r$, $i$, $z$, $J$, and $H$ bands are interpolated between $A_V$ and $A_K$. 

The estimated distances of \multicolor companions are reported in Table \ref{tab:AO_params}. For these companions with color information, \multicolorod have estimated distances that are 2-$\sigma$ inconsistent with the primary stars. Therefore, they are unlikely to be physically associated with the KOIs and thus are not considered in the following analyses. All \multicolorsubarc companions with color information and less than 1$^{\prime\prime}$ angular separations have consistent distances with their KOIs. This is consistent with the finding that stellar companions with sub-arcsec separations are mostly gravitationally bound to KOIs~\citep{Horch2014}. For stellar companions with 1.0$^{\prime\prime}$ to 3.0$^{\prime\prime}$ separations, 2 out of 11 (18\%) have inconsistent distances and are thus not physically associated with their KOIs. 

\subsection{Physical Association Based on Galactic Stellar Population Model}
\label{sec:phy_ass_galactic}

For the stellar companions without color information, we cannot adopt the method described in \S \ref{sec:phy_ass_color}. However, their physical association needs to be addressed because the frequency of optical doubles/multiples is not negligible: \multicolorod out of \multicolor stellar companions with color information are not gravitationally bound. We therefore develop a statistical approach to assess the physical association of detected stellar companions. 

Using the TRILEGAL galaxy model~\citep{Girardi2005}, we run two sets of simulations. In the first set, we turn off binary parameters and calculate the fraction of optical doubles/multiples as a function of $K_1$, $K_2$ and $\Delta\theta$, where $K_1$ is the magnitude of primary star, $K_2$ is the magnitude of the brightest nearby star, $\Delta\theta$ is the radius range in arcsec. In the second set of simulation, we consider both optical doubles/multiples and gravitationally bound systems. From results of both sets of simulation, we can calculate the relative contribution of optical doubles/multiples and gravitationally bound stellar systems at a given combination of $K_1$, $K_2$ and $\Delta\theta$, which allows us to calculate the probability of physical association in the absence of color information. 

In each simulation, ten fields with a FOV of 1 square degree are simulated. These fields have different galactic latitudes so the combination of the results from the fields gives a better statistical result of the entire Kepler FOV. We consider two different filters, $J$ and $K$ bands because all detections in single filter are in either $J$ or $K$ band. The majority (\singleK out of \singleall) are in $K$ band. The physical association probabilities of detected stellar companions in single filter are given in Table \ref{tab:AO_params}. We also provide a calculator for the probability of physical association as a function of $K_1$, $K_2$ and $\Delta\theta$ in $r$, $z$, $J$, $H$ and $K$ filters\footnote{http://www.astro.yale.edu/jwang/Cal\_Prob\_PA.py}. 

\subsection{Comparing Two Physical Association Methods}
\label{sec:phy_ass_comp}

We check the consistency of two methods for testing physical association. Since there are \multicolor stellar companions with color information, we can use this sample to perform the test: how physical association probabilities (in $K$ band) correlate with acceptances/rejections based on color information. We divide the  physical association probabilities into 3 intervals, [0.00-0.33], [0.33-0.67], and [0.67,1.00]. For the lower probability interval [0.00-0.33], 2 out of 3 stellar companions (KOI-98 and KOI-377) are rejected based on color information at 2-$\sigma$ level. For the median probability interval [0.33-0.67], 2 out of 3 stellar companions (KOI-377 and KOI-3444) are rejected based on color information. For the higher probability interval [0.67-1.00], 2 out of 15 stellar companions (KOI-97 and KOI-1812) are rejected based on color information. These results demonstrate consistency between these two methods. At a physical association probability smaller than 0.33, despite small number statistics, the majority (67\%) of stellar companions are rejected based on color information. In contrast, at a physical association probability higher than 0.67, the majority of stellar companions (87\%) show consistent colors to be physically associated. 

\section{Synthesizing AO Observations with Other Techniques}
\label{sec:aorvda}

While \detectstar stellar companions around \detectsys KOIs are detected via AO observations, the AO technique is not sensitive to stellar companions that are too close to spatially resolve, nor is it sensitive to stellar companions that are too faint to detect with a sufficient SNR. By conducting simulations, we can calculate the search completeness of AO observations. 

We define a parameter space, $a-i$ space, where $a$ is the semi-major axis of a companion star, and $i$ is the angle between the sky plane and the companion star orbital plane. We divide the parameter space into a grid ($\Delta a=0.5$ AU, $\Delta i=10^\circ$). We simulate 1000 companion stars at each gridpoint in the $a-i$ parameter space. The mass ratio distribution of simulated companions follows a Gaussian distribution from~\citet{Duquennoy1991}, i.e., $\overline{q} = m_2/m_1 = 0.23$, $\sigma_q = 0.42$. We use the median orbital eccentricity for binary stars ($e=0.4$) and a random true anomaly distribution in simulations. If the contrast ratio ($\Delta$ Mag) between a simulated companion and the primary star is smaller than the value given by the 5-$\sigma$ AO contrast curve, then we record it as a detection. The median AO completeness contours are plotted in Fig. \ref{fig:RV_AO_DA_completeness}.

For the parameter space on the $a-i$ plane that AO is not sensitive to, we use other observations or techniques to constrain the presence of stellar companions. 

\subsection{Radial Velocity Observation}
\label{sec:rv}

There are \rvstar KOIs in our sample with at least 3 epochs of RV observation. Following the description of ~\citet{Wang2014b}, we use the Keplerian Fitting Made Easy (KFME) package~\citep{Giguere2012} to analyze the RV data. For cases in which the number of RV data points are not adequate to constrain a Keplerian orbit, we use linear fitting to check if the RV data exhibit long-term trend. The RV data serve two purposes. First, they reveal stellar companions via RV trends. Among \rvstar KOIs with RV data, however, only KOI-5 exhibits a RV trend. The stellar companion that can potentially induce the trend is constrained to be beyond 7 AU~\citep{Wang2014b}. More recent RV data suggest that, in addition to two transiting planet candidates, two more distant components exist in KOI-5 system (Howard Isaacson, private communication). One is a sub-stellar companion with a period of $\sim$2700 days and the other one is the AO-imaged stellar companion. Therefore, we consider the closest stellar companion to KOI-5 to have a projected separation of 40.3 AU (Table \ref{tab:AO_params}).

The second purpose the RV data serve is to constrain the presence of stellar companions in the non-detection cases. Given the RV data, we can study the completeness of searching for stellar companions by simulations~\citep{Wang2014a,Wang2014b}. Similar to the AO completeness study, we simulate 1000 companion stars on each grid point and count the number of simulated companion stars that can be detected given the time baseline, observation epochs, and measurement uncertainties of the RV data. The median RV completeness contours are plotted in Fig. \ref{fig:RV_AO_DA_completeness}.

\subsection{Dynamical Analysis}
\label{sec:dynamical}
In addition to the RV and AO data, further constraints on potential stellar companions can be placed on multi-planet systems. There are \totalmulti (32\% of the sample) multi-planet systems in our sample for which we can apply a dynamical analysis~\citep{Wang2014a}. This dynamical analysis makes use of the co-planarity of multi-planet systems discovered by the Kepler mission~\citep{Lissauer2011}. A stellar companion with high mutual inclination to the planetary orbits would have perturbed the orbits and significantly reduced the co-planarity of planetary orbits, and hence the probability of multi-planet transits. Therefore, the fact that we have observed multiple transiting planet helps to exclude the possibility of a highly-inclined stellar companion. The dynamical analysis is complementary to the RV technique because it is sensitive to stellar companions with large mutual inclinations to the planetary orbits. The parameter space to which the dynamical analysis is sensitive is shown in Fig. \ref{fig:RV_AO_DA_completeness}.  

\subsection{Combining Results From Different Techniques}
\label{sec:comb}

For the RV and AO observations, detection completeness contours are calculated based on simulations given the time baseline, cadence, measurement uncertainties, and the contrast curve. For the dynamical analysis, numerical integrations give the fraction of time when multiple planets can stay with small mutual inclinations ($<5^\circ$) so that multiple transiting planets can be observed~\citep{Wang2014a}. We denote $c_{\rm{RV}}$, $c_{\rm{AO}}$ and $c_{\rm{DA}}$ as the completenesses at a given point in the $a-i$ parameter space, overall completeness $c$ is equal to $1-(1-c_{\rm{RV}})\times(1-c_{\rm{AO}})\times(1-c_{\rm{DA}})$. 

The completeness is then integrated over the $a-i$ parameter space. {For the integration, distribution functions of $a$ and $i$ are necessary to account for contribution at different places in $a-i$ parameter space. The result of the integration is sensitive to the adopted distribution function. Since the distribution function of $a$ is uncertain for plant host stars and measuring the distribution is the main goal of this paper, we adopt an iterative approach to incorporate $a$ distribution of stellar companions. For the first iteration, we assume a log-normal distribution for $a$~\citep{Duquennoy1991, Raghavan2010}. However, this distribution is not representative for stars with planets~\citep{Wang2014b}, so in the subsequent iterations we adopt the $a$ distribution from \S \ref{sec:planet_frequency}. The iteration stops when $a$ distributions from two consecutive iterations differ less than 1\% at any separations. 

We assume a random distribution of $-\cos i$ for systems with only one transiting planet, and the $i$ distribution from~\citet{Hale1994} for systems with multiple transiting planets. The treatment for multiple transiting planet systems is detailed in~\citet{Wang2014a}, i.e., a coplanar distribution for stellar companions within 15 AU, a random $-\cos i$ distribution for stellar companions beyond 30 AU, and a mixture  of the previous two $i$ distributions for intermediate separations between 15 and 30 AU. 

\subsection{Correcting For Detection Bias Against Planets in Multiple-Star Systems}
\label{sec:bias}

Planets in MSS are more difficult to find using the transit method because of flux contamination. The effect of this bias and a correction method have been discussed in~\citet{Wang2014b}. We briefly introduce the method here. 

We conduct simulations to quantify the detection bias against planets in MSS. For each KOI, we choose the one planet that gives the highest SNR. We add a companion star in the system and calculate the SNR in the presence of flux contamination for two cases: planet transiting the primary star and planet transiting the secondary star. If the SNR is higher than 7.1~\citep{Jenkins2010}, then the planet can still be detected, but with a lower significance. We randomly assign a stellar companion (secondary star) to a KOI (primary star) and repeat this procedure 1000 times for both the primary and the secondary star. We record the fraction of planet detections in 2000 simulations considering flux contamination. We designate the fraction to be $\alpha$, which will be used in correcting for the bias of detecting planets in MSS. For example, $\alpha=0.95$ indicates that 95\% of planets would still be detected in the presence of flux contamination. In order to account for the 5\% missed planets, for every $N$ MSS that host such a planet, we should use $N/\alpha$ to represent the underlying MSS population that host such a planet. Since the transiting signal of gas giant planets is large, they are rarely missed in Kepler observations. Therefore, $\alpha$ is close to one in most cases. 

\section{Stellar Multiplicity Rate For Kepler Stars with Gas Giant Planets}
\label{sec:planet_frequency}

The Kepler mission has provided us with a large sample of planet candidates. However, we do not know $a\ priori$ whether a given planet host star is in SSS or MSS. Follow-up observations are critical in identifying additional stellar companions in planetary systems. Even in the case of non-detection with RV and AO, we can calculate the probability of a star being in a MSS based on the completeness study (\S \ref{sec:comb}). For example, given the overall completeness $c$ and the stellar multiplicity rate (MR), the probability of the star having an undetected companion (or being in a MSS) within $r$ (in AU) is: 
\begin{equation} 
\label{eq:nmns1}
p_M(a\leq r)=\int_{a\leq r}{\rm{MR}}(a)\int(1-c(a,i))\omega(i)\ d i\ d a,
\end{equation}
where $\omega(i)$ is a weighting function for $i$. For single planetary systems, $\omega(i)di=d(-\cos i)$. For multiple planetary systems, $\omega(i)di$ is a piecewise function depending on stellar separation $a$ (\S \ref{sec:comb}). The form of the weighting function for $a$, MR($a$), was also discussed in \S \ref{sec:comb}. MR($a$) is the $a$ distribution of stellar companions for planet host stars. Here, we use MR($a$) as a differential distribution, which is the derivative of a cumulative distribution MR($a\leq r$), i.e., Equation \ref{eq:mfpl}, where both $a$ and $r$ are semi-major axis of an orbit. MR($a$) and MR($a\leq r$) are derived in an iterative way. For each iteration, we use MR($a$) from the previous iteration in Equation \ref{eq:nmns1} to calculate $p_M$, which is then fed into Equation \ref{eq:nmns} and \ref{eq:mfpl} to calculate MR($a$) in the new iteration. The iteration converges until MR($a$) from the new iteration and MR($a$) from the previous iteration agree within 1\% at any separations. Following this procedure, we calculate the number of MSS, $N_M$, and the number of SSS, $N_S$. Since $N_M$ and $N_S$ are the sums of probabilities, they are not necessarily integers:
\begin{equation} 
\label{eq:nmns}
N_M(a\leq r)=\sum\limits_{k=1}^{N} [p_M(a\leq r,k)/\alpha(k)],\ N_S(a\leq r)=\sum\limits_{k=1}^{N} [1-p_M(a\leq r,k)],
\end{equation}
where $N$ is the total number of stars in the sample, $p_M(k)$ is the probability of the $k_{\rm{th}}$ star being in a MSS, $\alpha(k)$ is the correction factor for the detection bias for planets in MSS (discussed in \S \ref{sec:bias}). Note that there is an implicit correction factor for SSS in Equation \ref{eq:nmns}, but that this factor is 1. If a physically associated stellar companion is detected within a semi-major axis $r$ to a KOI, then $p_M(a\leq r)$ is assigned to 1. We note that AO observation only measures projected separation. The conversion from projected separation to semi-major axis is addressed by a Monte-Carlo simulation assuming that stellar companions have randomly oriented orbits (\S \ref{sec:pap}). We also assign $\alpha$ to 1 because no bias exists in this case since a planet has already been detected in a MSS. The cumulative stellar multiplicity rate for planet host stars can be calculated:
\begin{equation} 
\label{eq:mfpl}
{\rm{MR}}(a\leq r)=\cfrac{N_M(a\leq r)}{N_M(a\leq r)+N_S(a\leq r)}.
\end{equation}

\subsection{Considering Physical Association Probability and Companion Orbital Orientation}
\label{sec:pap}

Not all AO detected stellar companions are physically associated with the KOI. Therefore, we need to consider the probability of physical association when calculating $N_M$, which is later used for the cumulative stellar multiplicity rate calculation (Equation \ref{eq:mfpl}). Similarly, $N_M$ may be different due to the orbit orientation of detected stellar companions. AO observation only measures projected separation, but we need semi-major axis in $N_M$ calculation. Since orbital orientation, eccentricity and true anomaly are required to covert projected separation to semi-major axis, and these are not known for a single epoch AO observation, the conversion cannot be performed on an individual system. However, we can run a Monte-Carlo simulation to calculate $N_M$ and its uncertainty due to physical association probability and companion orbital orientation. 

We developed two methods to calculate the probability of physical association in \S \ref{sec:phy_ass_color} and \S \ref{sec:phy_ass_galactic}. For detections in multiple filters, we estimate the distance of a stellar companion based on its color information. We exclude stellar companions whose distances are inconsistent with the KOI distance at more than 2-$\sigma$ level. For detections in only one filter, we estimate the probability of physical association using a galactic stellar population model . Then a random number following the uniform distribution between 0 and 1 is generated. If the random number is higher than the physical association probability, then the detection is excluded in the stellar multiplicity rate calculation. To account for the uncertainty in converting projected separation to semi-major axis, we assume randomly-orientated companion orbits. We use the median orbital eccentricity for companion stars ($e=0.4$) and a random true anomaly distribution in simulations. The calculation for $N_M$, $N_S$, and the stellar multiplicity rate is repeated for 1000 times for their values and uncertainties. 

\subsection{Treatments For Different Stellar Separations}
\label{sec:stfds}

For small separations, i.e., $a\leq$ 10 AU, the RV data provide an effective constraint on stellar companions. As shown in Fig. \ref{fig:RV_AO_DA_completeness}, the completeness of the RV technique is higher than 50\% for the majority of parameter space within 10 AU. However, RV data are available for only \rvstar out of \totaltargets KOIs. While considering all KOIs for stellar multiplicity rate within 10 AU seems to improve statistics, it in fact does not help because the majority of KOIs do not have data to constrain stellar companions within 10 AU. Instead, these KOIs without RV data outnumber KOIs with RV data and thus dominate the statistics. Since we use statistics of field stars in the solar neighborhood as an initial guess for stellar separation distribution for stellar companions (see \S \ref{sec:comb}), the lack of RV data for the majority of the KOI sample results in the lack of constraint for stellar companions within 10 AU. Therefore, the resulting stellar separation distribution for these \totaltargets KOIs would be similar to that of field stars in the solar neighborhood. The similarity of stellar separation distribution is not physical but rather a result of a lack of constraint from RV data for the majority of the sample. 

To avoid the above problem, we consider only KOIs with RV data when calculating stellar multiplicity rate for small separations. To define small separations, we choose separations at which the completeness of AO data becomes higher than the completeness of RV data. Based on Fig. \ref{fig:RV_AO_DA_completeness}, the transition separation is at 30-60 AU, so we adopt 50 AU as the transition separation. For stellar multiplicity rate within 50 AU, we consider \rvstar KOIs with both RV and AO data. For stellar multiplicity rate beyond 50 AU, we consider all \totaltargets KOIs for which we have AO data.  

One concern of using KOIs with RV data is the selection bias of RV observation and its potential influence on stellar multiplicity rate measurement. If RV observations are preferentially conducted for single stars or stars without significant flux contamination, then the stellar multiplicity rate for these KOIs would be lower because of selection bias. However, we have discussed this issue in Section 4.5 of~\citet{Wang2014b} showing no evidence of such selection bias. For this work, we also checked the 84 Kepler stars in our sample. For 19 stars that received RV follow-up observations, 5 have stellar companions within 2$^{\prime\prime}$, one has severe flux contamination, i.e., delta mag smaller than 2 mag. For 65 stars without RV data, 11 have stellar companions within 2$^{\prime\prime}$, 4 have severe flux contamination. The detection rates of stellar companions are comparable between stars receiving RV observations and stars without RV observations. Therefore, there is no evidence of selection bias of RV follow-up observations, i.e., stars with RV data tend to have fewer stellar companions or fewer bright companions than stars without RV data.

\subsection{Stellar Multiplicity Rate vs. Stellar Companion Separation}
\label{sec:mr_vs_a}

Fig. \ref{fig:Multi_Field} shows the comparison between the cumulative stellar multiplicity rate for field stars~\citep[blue hatched region,][]{Duquennoy1991,Raghavan2010} and that for planet host stars (red hatched regions). The field stars serve as a control sample for comparison. Hatched regions represents 1-$\sigma$ uncertainties. For field stars in the solar neighborhood, we adopt a 2\% uncertainty~\citep{Raghavan2010}. For planet host stars, we consider two sources of uncertainty. First, we consider the uncertainty induced by physical association (\S \ref{sec:pap}). Second, we consider Poisson noise by propagating the uncertainty in Equation \ref{eq:mfpl}. The two uncertainties are summed in quadrature for the final uncertainty. 

The stellar multiplicity rate for planet host stars is consistent with zero and stays flat at 0$^{+5}_{-0}$\% for stellar separations smaller than 20 AU. For 20 AU $<a<$ 200 AU, the stellar multiplicity rate is 34\%$\pm$8\%. The increase of the cumulative stellar multiplicity rate from [0-20] AU to [20-200] AU is much faster than the control sample whose cumulative stellar multiplicity rate changes from 18\% to 30\%. Beyond 200 AU, the stellar multiplicity rate for planet host stars is higher than that for the control sample, but they are consistent within measurement uncertainty. The slopes for the cumulative stellar multiplicity rate change are similar between planet host stars and the control sample. 


\section{Discussion}
\label{sec:discussion}

\subsection{Interpretation of the Stellar Multiplicity of Stars With Gas Giant Planets}
\label{sec:interpret}

According to Fig. \ref{fig:Multi_Field}, there may be three stellar separation ranges in which stellar companions affect gas giant giants formation and evolution differently. For stellar separations smaller than 20 AU, planet formation is suppressed. No stellar companion has been found within 20 AU for Kepler stars with gas giant planets. This leads to a zero stellar multiplicity (0$^{+5}_{-0}$\%) that is significantly lower than that for the control sample (18\%$\pm$2\%). 

For separations between 20 AU and 200 AU, we notice a drastic increase of the cumulative stellar multiplicity rate for planet host stars. In contrast to the stellar multiplicity rate of 12\%$\pm$2\% in this separation range for the control sample, the stellar multiplicity rate is 34\%$\pm$8\% for planet host stars. 
The higher stellar multiplicity rate for planet host stars suggests that stellar companions in this separation range play an important role in gas giant planet migration, e.g., via the Lidov-Kozai mechanism. Therefore, for 20 AU $<a<$ 200 AU, the role of stellar companions in planet migration is more important than their role in suppressing planet formation. However, there may be a caveat asserting the role of stellar companions in planet migration. The search completeness is $\sim$30-50\% between 20 and 200 AU (Fig. \ref{fig:RV_AO_DA_completeness}), which is the lowest in the $a-i$ parameter space. Therefore, the stellar multiplicity rate in this separation range is the most uncertain. 

For separation beyond 200 AU, the stellar multiplicity rate for planet host stars is comparable to that for the control sample, which indicates that gas giant planet formation and evolution is not significantly affected by a stellar companion. 

\subsection{Comparison to Stars with Small Planets}
\label{sec:comp_PaperII}

~\citet{Wang2014b} measured the cumulative stellar multiplicity rate for a sample of planet host stars. This sample is dominated by stars with smaller planets, 43 out 56 stars have only transiting planets that are smaller than 3.8 $R_\oplus$. We compare the stellar multiplicity rates from ~\citet{Wang2014b}  and this work, which focuses on gas giant planet host stars. While they are qualitatively identical, the characteristic separations are different. For example, the stellar multiplicity rate for small planet host stars intersects with control sample at $\sim$1500 AU whereas the intersection takes place at $\sim$100 AU for gas giant planets. The effective range of a stellar companion is an order of magnitude larger for small planets than for large planets: smaller planets are more prone to the influence of a stellar companion. 

There may be several explanations. First, in a multi-planet system, the timescale for pericenter precession and nodal precession increases with decreasing planet mass~\citep{Takeda2008}. For planet systems of the same orbital configurations, the Kozai timescale is more likely to be shorter than the timescale for pericenter precession and nodal precession for smaller planets, which makes these systems dominated by the Kozai effect due to the stellar companion. Therefore, smaller planets are more prone to the influence of a distant stellar companion because of a weaker planet-planet dynamical coupling. 
Second, for planet systems with both small and large planets, dynamical interaction between these two types of planets tend to eject more small planets than large planets (Xie et al. 2015 in prep.). In the presence of stellar companions, the dynamical interaction becomes more frequent and thus leads to higher loss of small planets. 

\subsection{Correlation Between Stellar Companion Properties and Planet Properties}
\label{sec:mr_vs_per}

Planet formation is subject to the influence of stellar companions. Therefore, planets with different properties (e.g., orbital period and planet radius) may be a result of different properties of stellar companions. Here, we study whether the difference of stellar companions results in the difference of planet properties. 

For properties of stellar companions, we focus on their 
differential magnitudes ($\Delta$ Mag). Since the majority (\starK out of \detectstar) of detected stellar companions have differential magnitudes in $K$ band, we use the differential magnitudes in $K$ band. For those whose $K$ band differential magnitudes are not available, we do not use them in the following analysis.

To find evidence that stellar companion properties affect planet properties, we adopt a K-S-test-based method that has been used to search for different populations divided by a parameter~\citep{Quinn2014,Buchhave2014}. The method is described as follows. We choose a value $x$ for a planet property parameter. The value divides the sample into two sub-samples. We then compare the stellar companion properties of two sub-samples with the K-S test. The $p$ value of the K-S test is recorded. By varying $x$ and repeating the K-S test, we obtain a function $p(x)$. If there are certain $x$ values that result in significant difference between two sub-samples, these values may represent characteristic values that divide different planet populations. These populations are a result of different properties of stellar companions. Because not all stellar companions are physically associated, we need to account for the effect of inclusion of optical double/multiples. The uncertainty due to this effect is addressed in a Monto Carlo simulation. In each trial, we draw a subset of detected stellar companions based on the probability of their physical association. We apply the K-S-test-based method to the subset and record $p(x)$. We repeat the process for 1000 times and find the median and confidence intervals at different levels. 

Fig. \ref{fig:SlidingKS} shows the $p$ value in K-S test as a function of planet orbital period. The K-S test compares the $\Delta$ Mag of two sub-samples divided by orbital period. The K-S test-based method suggests that there may be three different populations of gas giant planets. We caution that this finding is not conclusive at this stage given small number statistics. If using p=0.05 as a threshold, the dividing orbital periods are $\sim$10 and $\sim$70 days. The p-value dip at P$\sim$10 day is broad extending from 7 to 22 days. Hereafter we use P$\sim$10 days as a representative value. The stellar properties of these three populations may be distinctively different. Fig. \ref{fig:DeltaMagDist} shows the $\Delta$ Mag distribution. Stars with gas giant planets with $P<10$ days tend to have more stellar companions with small differential magnitudes ($\Delta$ Mag $<$ 2). In comparison, $\Delta$ Mag for stars with gas giant planets with $P\geq70$ days tend to be fainter but peak at $\Delta$ Mag $\sim$ 3. However, the peak may be due to a lack of sensitivity for fainter stellar companions. The $\Delta$ Mag for Stars with gas giant planets with $10\leq P<70$ days lies in between the previous two populations. 

We check whether the finding of three distinct populations is prone to inclusion of false positives. One major cause of false positive is underestimation of radius due to flux contamination. ~\citet{Horch2014} estimate the effect as a function of  $\Delta$ Mag in Kepler band in two scenarios: object transiting primary star and secondary star. After applying their calculation, there are five KOIs that could be false positives. All the potential false positives are for the scenarios in which an object is transiting secondary star. These KOIs are KOI-17, KOI-375, KOI-633, KOI-2672, and KOI-3678. Among them, KOI-17, also known as Kepler-6, has been confirmed~\citep{Dunham2010}. KOI-2672 is a system with multiple transiting planet candidates, so the false positive probability is low~\citep{Lissauer2012}. The stellar companions for KOI-375, KOI-633, and KOI-3678 are only detected in $K$ band and have differential magnitudes of 3.3, 3.9 and 3.3 mag. If physically associated, the $K$ band differential magnitudes would put the stellar companions in the range of early-type M dwarfs. Such low-mass stars are less likely to harbor gas giant planets, which makes the scenario in which an object transits the secondary star less probable. Given the false positive rate of $\sim$20\%~\citep{Fressin2013}, there would be at most one KOI that is a false positive, which will not significantly reduce the peak seen in Fig. \ref{fig:DeltaMagDist}. However, we must caution again that the three-population hypothesis is preliminary and remains to stand the test of more observations and a larger sample. 

\subsection{Comparison to Other Results}
\label{sec:comp}

The stellar multiplicity rate of planet host stars has been a research interest for a number of works (see \S \ref{sec:intro} for references). All previous results concerning the Kepler sample were for small planet host stars. This is because Kepler discoveries are dominated by small planets and these planets are of great interest for potential habitable worlds.~\citet{Ngo2014} used a sample of stars with gas giant planets that were discovered by ground-based RV and transiting surveys. In their ``Friends of Hot Jupiters" project, they found that the stellar multiplicity for 50 stars with hot Jupiters (HJs) is $48\pm9\%$ for the stellar separation range of 50-2000 AU. We have 28 HJs in our sample, the stellar multiplicity rate we find for these separations is $25\pm20\%$. However, if considering separations smaller than 50 AU, the stellar multiplicity rate within 2000 AU is $51\pm13\%$. The significant difference is due to the drastic change of stellar multiplicity rate between 20 and 200 AU (Fig. \ref{fig:Multi_Field}).  

In \S \ref{sec:mr_vs_per}, we suggest that the stellar companions around HJ host stars are on average brighter than stellar companions around stars hosting longer-period gas giant planets (Fig. \ref{fig:DeltaMagDist}). In our sample, all stellar companions around HJ host stars have $K$ band differential magnitudes smaller than 2.5 with one exception of KOI-17 ($\Delta$ Mag = 3.8 mag). The trend is absent in~\citet{Ngo2014}. None of their detected stellar companions have differential magnitudes smaller than 2.5 mag. One explanation is that they focus on stars with HJs discovered by ground-based RV and transiting surveys. In these surveys, a strong bias against MSS exists in the target selection and follow-up observations. The fact that ~\citet{Ngo2014} detected many fainter stellar companions suggests that there may be a missing population of HJ systems in equal-mass binary systems. Recent discovery of HJs in WASP-94 AB system is one example of such systems~\citep{Neveu2014}.

~\citet{Law2014} observed 715 KOIs using the Robo-AO system, they found that stars hosting short-period ($P\leq15$ days) giant planets are 2-3 times more likely to have stellar companions than their longer-period counterparts. For our sample, we have 32 giant planets with period shorter than 15 days. When comparing the stellar multiplicity rate for stars hosting short-period planets and the rest of our sample, we do not find a significant difference. In fact, the stellar multiplicity rates for the stars with short-period planets, stars with planets with $P>15$ days, and the entire sample are all $\sim$50\% within 5000 AU. For stars with short-period planets, they tend to have brighter stellar companions, which are easier for Robo-AO to detect. For stars with planets with longer periods, their stellar companions tend to be fainter with differential magnitudes peaking at $\Delta$ Mag = 3.0 in $K$ band. These faint stellar companions are 1-2 mag fainter in the visible bands at which Robo-AO operates. Thus they are likely to be missed by Robo-AO given the detection limits are $\sim$4-5 mag in $r$ and $i$ band for the median performance. The portion of fainter stellar companions missing in the Robo-AO survey may explain their finding that stars hosting short-period ($P\leq15$ days) giant planets are 2-3 times more likely to have stellar companions. Therefore, we suspect that the finding in~\citet{Law2014}  may be caused by a lack of sensitivity to faint stellar companions. 

~\citet{Jang2015} quantified planet formation efficiency in close binaries. The paper gave an empirical equation of gas giant planet formation efficiency as a function of primary star mass, mass ratio, binary separation and binary orbital eccentricity. Our finding of a suppressive planet formation within 20 AU is consistent with the conclusion in ~\citet{Jang2015}. However, ~\citet{Jang2015} predicted that gas giant planets in close binaries with separations smaller than 20 AU are not entirely impossible. In comparison, there is no such planetary system in our sample. The closest stellar companion we have detected is at 40.3 AU projected separation. There are several other planetary systems in close binaries detected by the RV technique, such as $\gamma$ Cep and HD 41004, but none of them has a binary separation smaller than 20 AU. The contrast between observations and theoretical predictions implies that a close stellar companion not only decreases the planet formation efficiency but also negatively influences planet evolution. ~\citet{Jang2015} also discussed the effect of mass ratio on planet formation. While we found the tentative correlation between hot Jupiter occurrence and equal-brightness stellar companion, it is difficult to make a direct comparison because ~\citet{Jang2015} focused on stellar companions with separations smaller than 100 AU whereas all but one companions in our study have separations larger than 100 AU. 

\section{Summary}
\label{sec:Summary}

We select a sample \totaltargets KOIs with \totalplanets gas giant planets to study the influence of stellar companions on planet formation. We obtain AO images for \myaostar KOIs with two instruments: PHARO at Palomar 200-inch telescope and NIRC2 at Keck II telescope. For the rest of the sample, we use AO images available from the CFOP database. In total, we have detected \detectstar stellar companions around \detectsys KOIs. 

Since some of the detected stellar companions are not physically associated with the KOIs, we develop two methods of testing the physical association. The first method makes use of the color information. For stellar companions with detections in multiple filters, we estimate their masses and absolute magnitudes based on their colors. Their distances can be estimated from the absolute magnitudes, apparent magnitudes, and extinctions. By comparing the distances of stellar companions and the KOIs, we can test their physical association. The second method is based on the statistics of galactic stellar population. For stellar companions that are detected in only one filter, the color information is missing. Instead, we run galactic stellar population model to simulate the Kepler field. With the results, we estimate the relative probability of a gravitationally bound stellar companions and optical doubles/multiples. Then we can calculate the probability of physical association as a function of angular separation and differential magnitude of a stellar companion. With these two methods of testing physical association, we can effectively exclude the effect of foreground and background stars on stellar multiplicity rate. 


We find that the stellar multiplicity rate for planet host stars is 0$^{+5}_{-0}$\% within 20 AU. In comparison, the stellar multiplicity rate is 18\%$\pm$2\% for the control sample, i.e., field stars in the solar neighborhood. The deficiency of stellar companions for planet host stars indicates that gas giant planet formation is suppressed by stellar companions within 20 AU. The stellar multiplicity rate for planet host star is 34\%$\pm$8\% for separations between 20 and 200 AU, which is higher than the control sample at 12\%$\pm$2\%. This suggests that stellar companions in this separation range play a important role in gas giant planet migration. Beyond 200 AU, stellar multiplicity rates are comparable between planet host stars and the control sample.

We explore whether stellar companions of different properties lead to different planet properties. We find evidence of three distinct populations of gas giant planets. They are separated by two characteristic orbital periods, 10 and 70 days. The stellar companions around stars with each planet population have different differential magnitude distributions. For example, stars with HJs ($P<10$ days) tend to have bright stellar companions with $K$ band differential magnitudes smaller than 2 mag. Stars with gas giant planets with $P\geq70$ days tend to have stellar companions whose differential magnitudes peak at $\sim$3.0 mag in $K$ band, which correspond to early type M dwarf companions. We emphasize that the three-population hypothesis is still tentative, and more follow-up observations are needed to either support or disprove it. 

If the three-population hypothesis survives future tests, it may have significant implications on gas giant planet formation and evolution. First, the migration of gas giant planets may be dependent upon the mass of a stellar companion. The majority of gas giant planets in our sample cannot form in situ because of a lack of building and accreting materials. They must migrate in via some mechanisms. The  migration mechanisms could be the same for different planet populations, e.g., the Lidov-Kozai mechanism. Or the migration mechanisms could be different, e.g., the Lidov-Kozai mechanism for HJs and in-disk migration for longer-period planets. In either case, the migration mechanisms have to be consistent with the observed mass-dependency of stellar companions. 

Second, if HJs tend to be preferentially found in binary star systems with similar magnitudes, this population of HJs have been neglected because of the concern of flux contamination. The HJs in WASP-94 AB~\citep{Neveu2014} and Kepler-13 AB~\citep{Szabo2011,Mazeh2012,Shporer2014,Johnson2014} systems may be the tip of iceberg for this population. Surveys that do not bias against MSS can constrain the HJ occurrence rate in binary star systems with similar magnitudes. Current and future space missions for transiting planets, e.g., the K2 mission~\citep{Howell2014} and the TESS mission~\citep{Ricker2014}, may serve the purpose. In fact, the transiting signal of HJs in binary stars may already exist in ground-based transiting surveys. In addition, AO-fed spectrographs will play an important role in validating or confirming HJs in binary stars~\citep[e.g., ][]{Crepp2014}.

\noindent{\it Acknowledgements} The authors would like to thank the anonymous referee whose comments and suggestions greatly improve the paper. We are grateful to telescope operators and supporting astronomers at the Palomar Observatory and the Keck Observatory. Some of the data presented herein were obtained at the W.M. Keck Observatory, which is operated as a scientific partnership among the California Institute of Technology, the University of California and the National Aeronautics and Space Administration. The Observatory was made possible by the generous financial support of the W.M. Keck Foundation. The research is made possible by the data from the Kepler Community Follow-up Observing Program (CFOP). The authors acknowledge all the CFOP users who uploaded the AO and RV data used in the paper. This research has made use of the NASA Exoplanet Archive, which is operated by the California Institute of Technology, under contract with the National Aeronautics and Space Administration under the Exoplanet Exploration Program. J.W.X. acknowledges support from the National Natural Science Foundation of China (Grant No. 11333002 and 11403012), the Key Development Program of Basic Research of China (973 program, Grant No. 2013CB834900) and the Foundation for the Author of National Excellent Doctoral Dissertation (FANEDD) of PR China.
J.W. acknowledges the travel fund from the Key Laboratory of Modern Astronomy and Astrophysics (Nanjing University). We thank Eric Ford for insightful comments on interpreting the stellar multiplicity rate of planet host stars.

\bibliography{mybib_JW_DF_PH5}

%
%
%
%

\begin{figure}[htp]
\begin{center}
\includegraphics[angle=0, width= 0.9\textwidth]{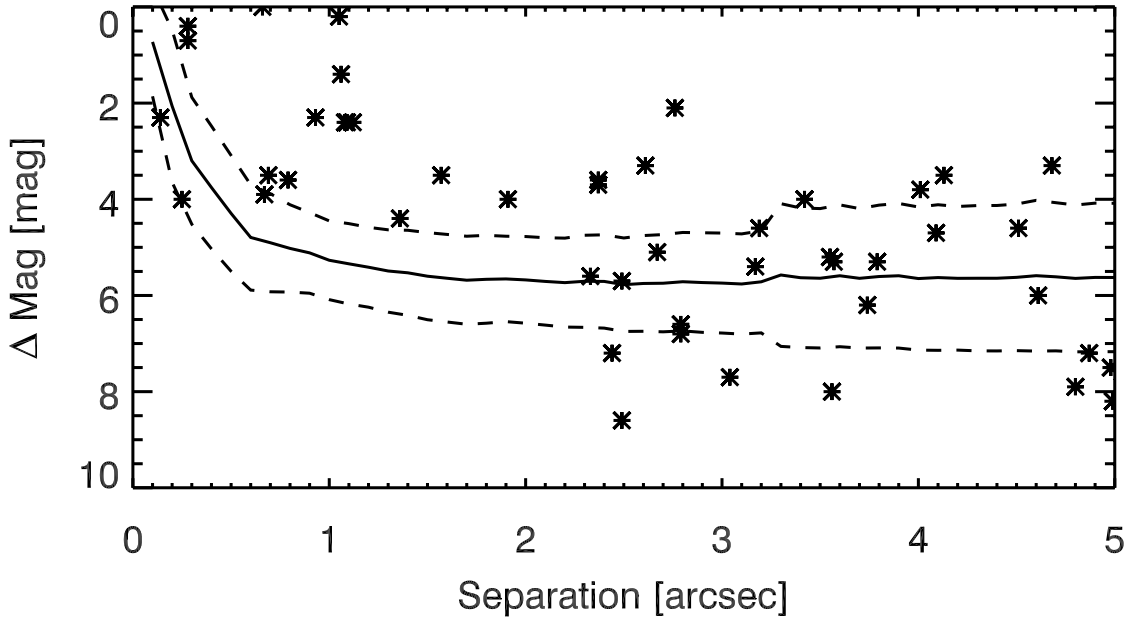} 
\caption{The median 5-$\sigma$ contrast curve for AO images of \totaltargets KOIs is shown in solid line. Dashed lines are 1-$\sigma$ deviation of the contrast curve. Detections within 5$^{\prime\prime}$ are shown as asterisks. When analyzing the detection completeness, each KOI is treated individually for the observation band in which the AO image was taken. A total of \detectstar visual companions around \detectsys KOIs are detected (Table \ref{tab:AO_params}). 
\label{fig:AO_contrast}}
\end{center}
\end{figure}
%
\begin{figure}[htp]
\begin{center}
\includegraphics[angle=0, width= 0.9\textwidth]{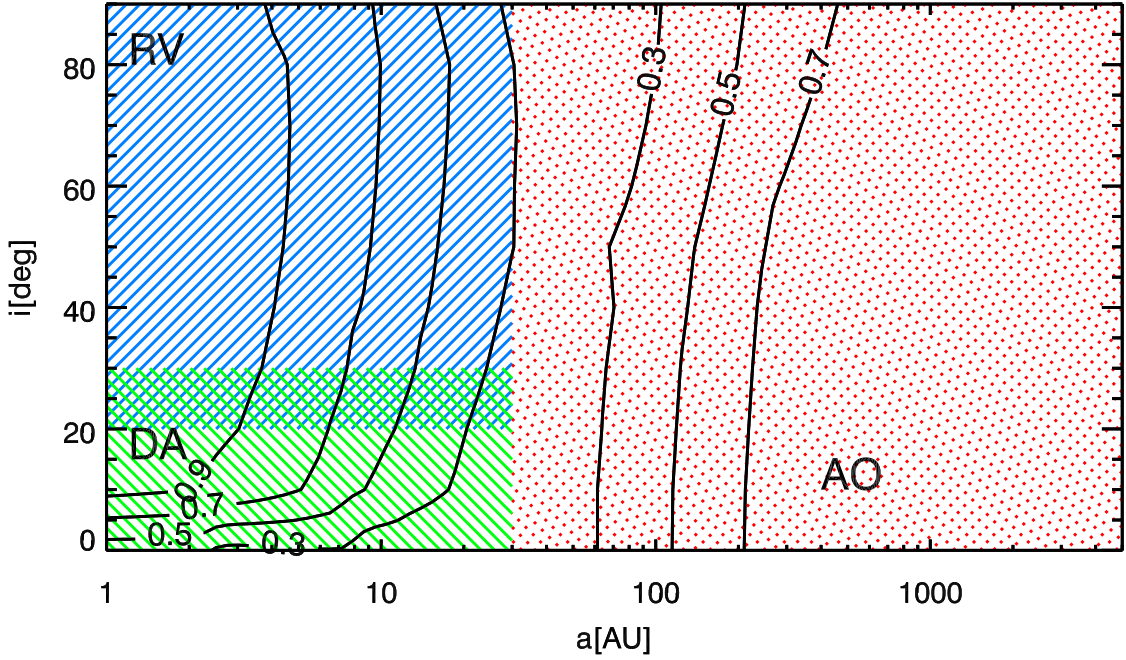} 
\caption{Typical completeness contours for 3 techniques used to detect and constrain stellar companions around planet host stars. The radial velocity (RV) technique is sensitive to stellar companions within $\sim$30 AU and with small or intermediate mutual inclinations to planet orbital planes (blue hatched region). Note that $i$ is the angle between the stellar companion orbital plane and the sky plane, so $i\sim90^\circ$ implies a small mutual inclination between the stellar companion orbital plane and a transiting planet orbital plane. Dynamical analysis (DA) is sensitive to stellar companions at larger mutual inclinations between the stellar companion orbital plane and a transiting planet orbital plane (green hatched region). The adaptive optics (AO) imaging technique is sensitive to stellar companions at wider orbits (red dotted region). The combination of these 3 techniques contributes to a survey of stellar companions with high completeness.   
\label{fig:RV_AO_DA_completeness}}
\end{center}
\end{figure}
%
\begin{figure}[htp]
\begin{center}
\includegraphics[angle=0, width= 0.9\textwidth]{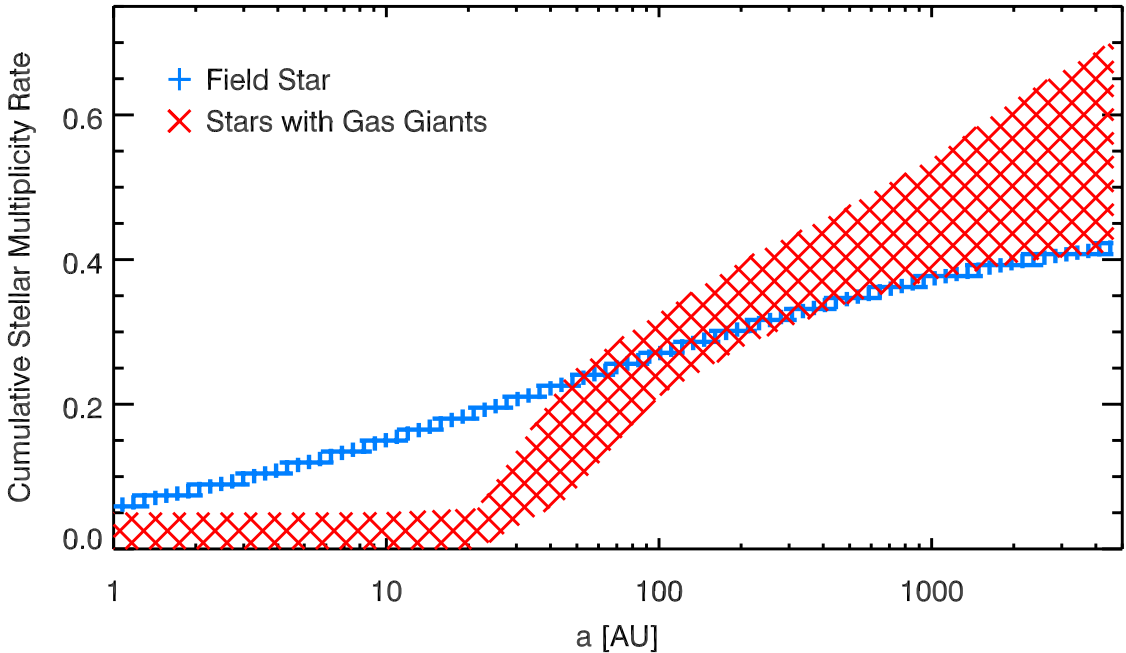} 
\caption{Comparison of the cumulative stellar multiplicity rate between field stars in the solar neighborhood (blue) and gas giant planet host stars (red). Hatched regions represent 1-$\sigma$ uncertainty regions. The stellar multiplicity rate for planet host stars is 0$^{+5}_{-0}$\% within 20 AU. In comparison, the stellar multiplicity rate is 18\%$\pm$2\% for the control sample. The stellar multiplicity rate for planet host star is 34\%$\pm$8\% for separations between 20 and 200 AU, which is higher than the control sample at 12\%$\pm$2\%. Beyond 200 AU, stellar multiplicity rates are comparable between planet host stars and the control sample     
\label{fig:Multi_Field}}
\end{center}
\end{figure}
%
\begin{figure}
\begin{center}
\includegraphics[angle=0, width= 0.9\textwidth]{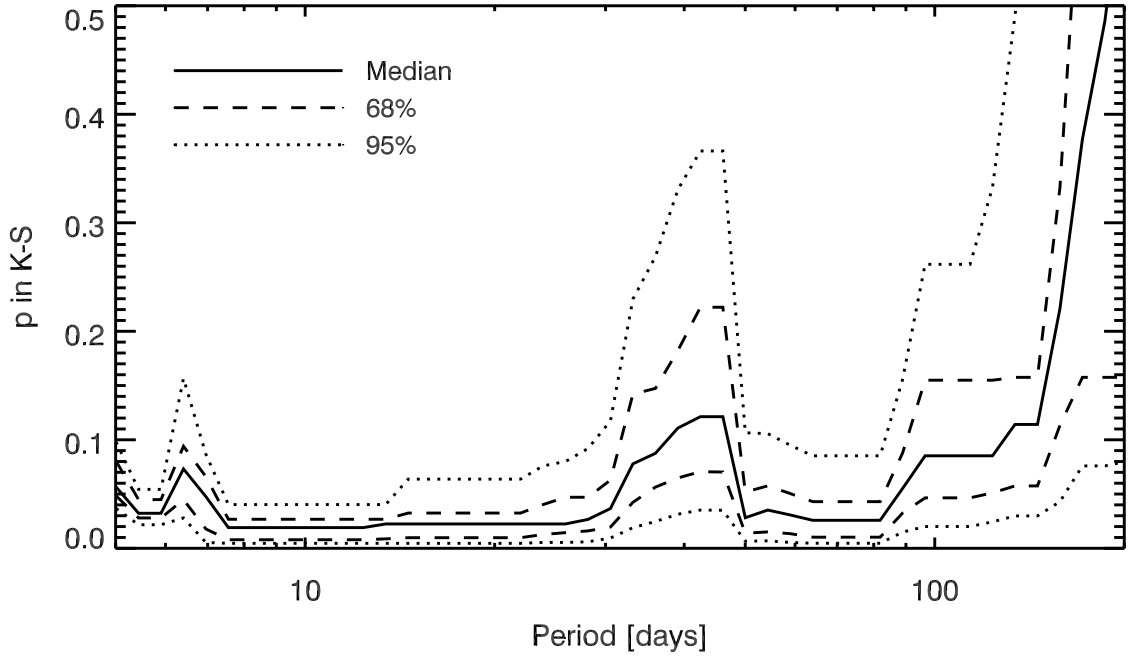} 
\caption{Results of K-S tests for the null hypothesis that stars with shorter-period and longer-period gas giant planets have stellar companions with similar distribution of differential magnitude. At each dividing period, we conduct a Monte Carlo simulation and show the median (solid), 68\% (dashed) and 95\% (dotted) confidence intervals. There may be three populations of gas giant planets whose formation and evolution are influenced differently by their stellar companions. The dividing periods for these populations are $\sim$10 and $\sim$70 days. 
\label{fig:SlidingKS}}
\end{center}
\end{figure}
%
\begin{figure}
\begin{center}
\includegraphics[angle=0, width= 0.9\textwidth]{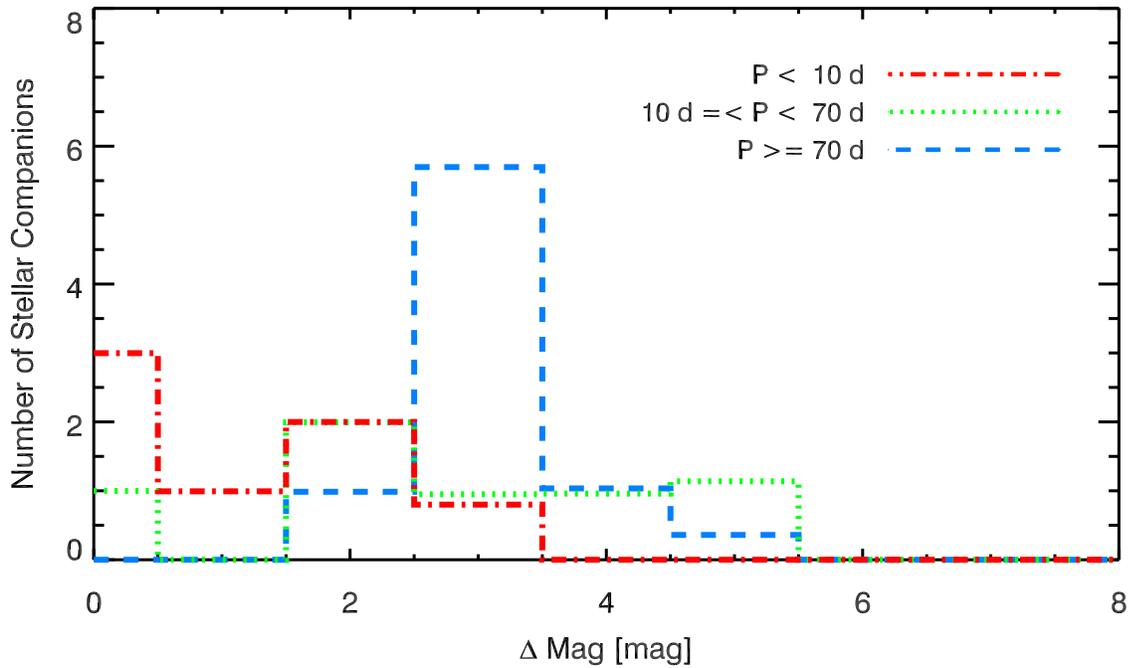} 
\caption{The distribution of differential magnitude for stellar companions. For stars with short-period gas giant planets ($P<10$ days), stellar companions tend to be bright with small differential magnitudes. For stars with gas giant planets with $P\geq70$ days, the differential magnitudes of their stellar companions tend to be fainter than the short-period counterparts. There is a peak at $\sim$3.0 mag, but it may be due to a lack of sensitivity for fainter stellar companions.  
\label{fig:DeltaMagDist}}
\end{center}
\end{figure}

\clearpage


\newpage
\begin{deluxetable}{lcccccccccccccllc}
\tablewidth{600pt}
\tabletypesize{\tiny}
\setlength{\tabcolsep}{0.01in}
\tablecaption{RV and AO data for \totalplanets KOIs\label{tab:stellar_params}}
\tablehead{
\multicolumn{11}{c}{\textbf{KOI}} &
\multicolumn{3}{c}{\textbf{RV}} &
\multicolumn{3}{c}{\textbf{AO}} \\
\colhead{\textbf{KOI}} &
\colhead{\textbf{KIC}} &
\colhead{\textbf{$\alpha$}} &
\colhead{\textbf{$\delta$}} &
\colhead{\textbf{K$_P$}} &
\colhead{\textbf{$T_{\rm eff}$}} &
\colhead{\textbf{$\log g$}} &
\colhead{\textbf{d}} &
\colhead{\textbf{\#PL}} &
\colhead{\textbf{R$_{\rm{PL}}$}} &
\colhead{\textbf{Period}} &
\colhead{\textbf{$T_{\rm start}$}} &
\colhead{\textbf{$T_{\rm end}$}} &
\colhead{\textbf{\#RV}} &
\colhead{\textbf{Telescope}} &
\colhead{\textbf{Band}} &
\colhead{\textbf{this work}} \\
\colhead{\textbf{}} &
\colhead{\textbf{}} &
\colhead{\textbf{(deg)}} &
\colhead{\textbf{(deg)}} &
\colhead{\textbf{(mag)}} &
\colhead{\textbf{(K)}} &
\colhead{\textbf{(cgs)}} &
\colhead{\textbf{(pc)}} &
\colhead{\textbf{}} &
\colhead{\textbf{(R$_{\oplus}$)}} &
\colhead{\textbf{(day)}} &
\colhead{\textbf{(MJD)}} &
\colhead{\textbf{(MJD)}} &
\colhead{\textbf{}} &
\colhead{\textbf{}} &
\colhead{\textbf{}} &
\colhead{\textbf{}} 
}

\startdata

K00001.01&11446443&286.808472&49.316399&11.338&5814.00&4.380&207.0&1&14.40$\pm$1.60&2.470613&54216.610352&56208.227539&18&Keck&JHK&\checkmark\\
K00003.01&10748390&297.709351&48.080853&9.174&4766.00&4.590&38.8&1&4.68$\pm$0.18&4.887800&54336.352539&56485.626953&42&WIYN&rz&\\
K00005.01&8554498&289.739716&44.647419&11.665&5861.00&4.190&286.6&2&5.66$\pm$0.72&4.780329&54983.515625&56486.439453&21&KeckPalomar&JK&\\
K00010.01&6922244&281.288116&42.451080&13.563&6025.00&4.110&919.7&1&15.90$\pm$2.10&3.522499&54983.540039&55781.534180&50&Palomar&J&\\
K00017.01&10874614&296.837250&48.239944&13.303&5826.00&4.420&494.3&1&11.07$\pm$0.54&3.234700&54984.560547&55043.519531&10&Palomar&J&\\
K00020.01&11804465&286.243439&50.040379&13.438&6011.00&4.230&608.4&1&17.60$\pm$2.50&4.437963&55014.412109&55761.325195&16&Palomar&J&\\
K00022.01&9631995&282.629669&46.323360&13.435&5972.00&4.410&593.4&1&11.27$\pm$0.70&7.891450&55014.403320&55792.438477&16&Palomar&J&\\
K00046.01&10905239&283.255493&48.355232&13.770&5764.00&4.400&604.1&2&4.33$\pm$0.37&3.487688&&&&Keck&K&\checkmark\\
K00063.01&11554435&289.226166&49.548199&11.582&5721.00&4.470&212.3&1&6.31$\pm$0.31&9.434158&&&&Keck&K&\checkmark\\
K00094.02&6462863&297.333069&41.891121&12.205&6184.00&4.181&463.6&4&4.22$\pm$0.42&10.423707&&&&MMT&JK&\\
K00094.03&6462863&297.333069&41.891121&12.205&6184.00&4.181&463.6&4&6.73$\pm$0.67&54.319930&&&&MMT&JK&\\
K00094.01&6462863&297.333069&41.891121&12.205&6184.00&4.181&463.6&4&11.40$\pm$1.10&22.343001&&&&MMT&JK&\\
K00097.01&5780885&288.581512&41.089809&12.885&5934.00&4.040&789.7&1&16.10$\pm$2.00&4.885489&55106.878906&55115.871094&9&MMT&JK&\\
K00098.01&10264660&287.708832&47.333050&12.128&6395.00&4.150&446.2&1&10.00$\pm$1.40&6.790123&55440.500987&56533.433594&7&MMTPalomar&JK&\\
K00108.02&4914423&288.984558&40.064529&12.287&5975.00&4.330&354.6&2&4.46$\pm$0.52&179.601000&55073.468750&56145.498047&22&KeckPalomar&JK&\\
K00119.01&9471974&294.559174&46.062328&12.654&5632.00&4.440&313.0&2&3.90$\pm$2.60&49.184310&&&&Palomar&JK&\checkmark\\
K00127.01&8359498&289.607971&44.345421&13.938&5731.00&4.450&617.4&1&10.93$\pm$0.47&3.578783&&&&Palomar&J&\\
K00128.01&11359879&296.200592&49.140121&13.758&5786.00&4.420&579.8&1&11.97$\pm$0.85&4.942783&55284.481445&55509.075195&30&Keck&K&\checkmark\\
K00131.01&7778437&299.097534&43.497589&13.797&6411.00&4.400&937.3&1&9.00$\pm$3.20&5.014233&&&&Keck&K&\checkmark\\
K00135.01&9818381&285.240845&46.668251&13.958&6082.00&4.370&810.1&1&10.56$\pm$0.46&3.024095&55752.008789&55806.833984&8&Keck&K&\checkmark\\
K00137.01&8644288&298.079437&44.746319&13.549&5385.00&4.430&421.9&3&4.75$\pm$0.43&7.641571&55075.508789&56146.484375&20&Palomar&J&\\
K00137.02&8644288&298.079437&44.746319&13.549&5385.00&4.430&421.9&3&6.01$\pm$0.54&14.858940&55075.508789&56146.484375&20&Palomar&J&\\
K00139.01&8559644&291.653168&44.688271&13.492&5952.00&4.380&612.0&2&7.34$\pm$0.66&224.797120&&&&Lick&J&\\
K00141.01&12105051&288.038300&50.651611&13.687&5425.00&4.500&463.1&1&5.43$\pm$0.29&2.624234&&&&MMT&JK&\\
K00152.01&8394721&300.517120&44.381580&13.914&6405.00&4.420&897.1&4&6.10$\pm$1.90&52.091100&&&&Palomar&K&\checkmark\\
K00157.03&6541920&297.115112&41.909142&13.709&5685.00&4.380&514.8&6&4.18$\pm$0.76&31.995467&55440.500987&56533.433594&7&Palomar&K&\checkmark\\
K00179.02&9663113&297.045410&46.328701&13.955&6081.00&4.420&771.2&2&5.00$\pm$1.70&572.397900&&&&Keck&K&\checkmark\\
K00244.01&4349452&286.638397&39.487881&10.734&6103.00&4.070&313.5&2&6.51$\pm$0.89&12.720365&55366.602539&56519.408203&104&KeckPalomar&JK&\\
K00279.01&12314973&295.486511&51.013500&11.684&6418.00&4.280&268.6&3&5.10$\pm$1.60&28.454899&&&&Keck&K&\\
K00289.02&10386922&282.945648&47.574905&12.747&5812.00&4.458&348.4&2&5.04$\pm$0.00&296.637100&56449.400391&56532.291016&5&KeckPalomar&K&\checkmark\\
K00318.01&8156120&288.153992&44.068821&12.211&6285.00&4.290&415.8&1&5.19$\pm$0.49&38.583360&&&&Palomar&K&\checkmark\\
K00319.01&8684730&290.117523&44.872940&12.711&5952.00&4.190&421.3&1&7.50$\pm$1.10&46.151590&&&&Palomar&K&\checkmark\\
K00340.01&10616571&297.664673&47.801392&13.057&5811.00&4.400&417.3&1&16.80$\pm$1.40&23.673188&&&&Palomar&K&\checkmark\\
K00344.01&11015108&283.340271&48.549042&13.400&5984.00&4.340&597.5&1&3.80$\pm$1.60&39.309240&&&&Palomar&K&\checkmark\\
K00351.02&11442793&284.433502&49.305161&13.804&6330.00&4.430&870.7&6&6.80$\pm$3.00&210.596590&&&&WIYN&rz&\\
K00351.01&11442793&284.433502&49.305161&13.804&6330.00&4.430&870.7&6&9.80$\pm$4.20&331.643000&&&&WIYN&rz&\\
K00366.01&3545478&291.664185&38.619255&11.714&6207.00&4.180&352.1&1&10.60$\pm$4.60&75.112019&&&&KeckPalomar&K&\checkmark\\
K00367.01&4815520&284.472168&39.911812&11.105&5569.00&4.360&153.6&1&4.98$\pm$0.71&31.578680&&&&KeckPalomar&JK&\checkmark\\
K00372.01&6471021&299.122437&41.866760&12.391&5872.00&4.487&355.2&1&8.52$\pm$0.23&125.630640&&&&MMT&K&\\
K00375.01&12356617&291.201202&51.144279&13.293&5757.00&4.140&778.2&1&10.40$\pm$1.40&988.881118&&&&Palomar&K&\checkmark\\
K00377.02&3323887&285.573975&38.400902&13.803&5777.00&4.450&619.4&3&8.22$\pm$0.59&38.907202&55342.448242&56506.363281&16&Palomar&JK&\\
K00377.01&3323887&285.573975&38.400902&13.803&5777.00&4.450&619.4&3&8.28$\pm$0.59&19.273938&55342.448242&56506.363281&16&Palomar&JK&\\
K00633.01&4841374&293.427032&39.942429&13.871&6070.00&4.030&640.1&1&5.80$\pm$1.80&161.479000&&&&Palomar&K&\checkmark\\
K00638.01&5113822&295.559418&40.236271&13.595&5980.00&4.310&575.9&2&3.80$\pm$1.20&23.636883&&&&MMT&K&\\
K00638.02&5113822&295.559418&40.236271&13.595&5980.00&4.310&575.9&2&3.90$\pm$1.30&67.093500&&&&MMT&K&\\
K00672.02&7115785&291.169525&42.640808&13.998&5524.00&4.410&524.6&3&4.03$\pm$0.41&41.749990&&&&Keck&K&\checkmark\\
K00680.01&7529266&292.287323&43.197281&13.643&6327.00&4.350&786.9&1&8.00$\pm$2.40&8.600145&&&&Keck&K&\checkmark\\
K00682.01&7619236&295.197998&43.269508&13.916&5592.00&4.250&632.7&1&10.80$\pm$1.40&562.142400&&&&Keck&K&\checkmark\\
K00683.01&7630229&297.824310&43.258381&13.714&5887.00&4.390&647.6&1&6.00$\pm$0.77&278.122200&&&&Palomar&K&\checkmark\\
K00686.01&7906882&296.840759&43.647121&13.579&5559.00&4.470&494.5&1&11.40$\pm$4.40&52.513565&&&&Palomar&K&\checkmark\\
K00697.01&8878187&289.000885&45.154270&13.684&5779.00&4.050&1315.7&1&4.24$\pm$0.75&3.032154&&&&Keck&JHK&\checkmark\\
K00707.03&9458613&289.077545&46.005219&13.988&5904.00&4.030&1236.7&5&4.10$\pm$0.36&31.784660&&&&Keck&K&\checkmark\\
K00707.02&9458613&289.077545&46.005219&13.988&5904.00&4.030&1236.7&5&4.69$\pm$0.41&41.027950&&&&Keck&K&\checkmark\\
K00707.01&9458613&289.077545&46.005219&13.988&5904.00&4.030&1236.7&5&5.50$\pm$0.48&21.775754&&&&Keck&K&\checkmark\\
K00716.01&9846348&297.492157&46.694519&13.754&6115.00&4.490&713.2&1&6.80$\pm$3.20&26.893084&&&&Palomar&K&\checkmark\\
K01162.01&10528068&288.868225&47.759430&12.783&6138.00&4.280&490.7&1&4.40$\pm$2.30&158.692400&&&&Palomar&K&\checkmark\\
K01206.01&3756801&293.954590&38.899971&13.642&5754.00&4.110&826.9&1&6.20$\pm$1.90&422.917678&&&&Keck&K&\checkmark\\
K01274.01&8800954&283.256805&45.087780&13.354&5310.00&4.550&356.9&1&4.73$\pm$0.28&704.962626&&&&Palomar&K&\checkmark\\
K01311.01&10713616&283.532959&48.094261&13.498&6188.00&4.200&648.2&1&4.20$\pm$1.70&83.577520&&&&Palomar&K&\checkmark\\
K01335.01&4155328&291.020142&39.220680&13.968&6222.00&4.040&1385.0&1&8.60$\pm$2.70&127.832900&&&&Palomar&K&\checkmark\\
K01353.02&7303287&297.465332&42.882839&13.956&6260.00&4.080&1094.5&2&3.80$\pm$1.10&34.543630&&&&Palomar&K&\checkmark\\
K01353.01&7303287&297.465332&42.882839&13.956&6260.00&4.080&1094.5&2&18.60$\pm$5.30&125.865460&&&&Palomar&K&\checkmark\\
K01375.01&6766634&288.320435&42.261414&13.709&6169.00&4.358&594.0&1&6.78$\pm$0.00&321.213900&&&&KeckPalomar&JHK&\checkmark\\
K01411.01&9425139&298.995087&45.909512&13.377&5753.00&4.410&502.0&1&7.05$\pm$0.84&305.056500&&&&Palomar&K&\checkmark\\
K01431.01&11075279&287.022278&48.681938&13.460&5649.00&4.460&486.8&1&8.45$\pm$0.48&345.161300&56472.390625&56532.501953&6&Palomar&K&\checkmark\\
K01439.01&11027624&290.851776&48.521339&12.849&5930.00&4.090&719.2&1&7.80$\pm$1.10&394.610700&55075.273438&56531.312500&6&Palomar&K&\checkmark\\
K01474.01&12365184&295.417877&51.184761&13.005&6293.00&4.270&552.0&1&9.30$\pm$1.20&69.732970&&&&Palomar&K&\checkmark\\
K01478.01&12403119&288.848633&51.209049&12.450&5493.00&4.417&288.4&1&5.26$\pm$0.35&76.133540&&&&Palomar&K&\checkmark\\
K01645.01&11045383&298.210938&48.559158&13.418&5197.00&4.530&379.9&1&10.31$\pm$3.10&41.166759&&&&Palomar&K&\checkmark\\
K01658.01&4570949&294.192108&39.618999&13.308&6422.00&4.320&750.6&1&12.35$\pm$0.67&1.544930&&&&Keck&K&\checkmark\\
K01684.01&6048024&293.534485&41.329899&12.849&6387.00&4.430&616.1&1&7.30$\pm$2.60&62.815570&&&&Palomar&K&\checkmark\\
K01779.02&9909735&298.482819&46.793621&13.297&5812.00&4.140&586.3&2&5.00$\pm$1.70&11.815018&&&&KeckPalomar&K&\checkmark\\
K01779.01&9909735&298.482819&46.793621&13.297&5812.00&4.140&586.3&2&5.80$\pm$1.90&4.662723&&&&KeckPalomar&K&\checkmark\\
K01783.02&10005758&289.341431&46.988239&13.929&6235.00&4.460&845.5&2&5.20$\pm$1.90&284.042300&&&&Palomar&K&\checkmark\\

\enddata

\end{deluxetable}

\begin{deluxetable}{lcccccccccccccllc}
\tablewidth{600pt}
\tabletypesize{\tiny}
\setlength{\tabcolsep}{0.01in}
\tablecaption{RV and AO data for \totalplanets KOIs\label{tab:stellar_params}}
\tablehead{
\multicolumn{11}{c}{\textbf{KOI}} &
\multicolumn{3}{c}{\textbf{RV}} &
\multicolumn{3}{c}{\textbf{AO}} \\
\colhead{\textbf{KOI}} &
\colhead{\textbf{KIC}} &
\colhead{\textbf{$\alpha$}} &
\colhead{\textbf{$\delta$}} &
\colhead{\textbf{K$_P$}} &
\colhead{\textbf{$T_{\rm eff}$}} &
\colhead{\textbf{$\log g$}} &
\colhead{\textbf{d}} &
\colhead{\textbf{\#PL}} &
\colhead{\textbf{R$_{\rm{PL}}$}} &
\colhead{\textbf{Period}} &
\colhead{\textbf{$T_{\rm start}$}} &
\colhead{\textbf{$T_{\rm end}$}} &
\colhead{\textbf{\#RV}} &
\colhead{\textbf{Telescope}} &
\colhead{\textbf{Band}} &
\colhead{\textbf{this work}} \\
\colhead{\textbf{}} &
\colhead{\textbf{}} &
\colhead{\textbf{(deg)}} &
\colhead{\textbf{(deg)}} &
\colhead{\textbf{(mag)}} &
\colhead{\textbf{(K)}} &
\colhead{\textbf{(cgs)}} &
\colhead{\textbf{(pc)}} &
\colhead{\textbf{}} &
\colhead{\textbf{(R$_{\oplus}$)}} &
\colhead{\textbf{(day)}} &
\colhead{\textbf{(MJD)}} &
\colhead{\textbf{(MJD)}} &
\colhead{\textbf{}} &
\colhead{\textbf{}} &
\colhead{\textbf{}} &
\colhead{\textbf{}} 
}

\startdata

K01783.01&10005758&289.341431&46.988239&13.929&6235.00&4.460&845.5&2&8.50$\pm$3.10&134.479720&&&&Palomar&K&\checkmark\\
K01784.01&10158418&297.646057&47.167488&13.592&5853.00&4.540&574.0&1&5.20$\pm$3.00&5.007410&&&&Keck&JHK&\checkmark\\
K01792.01&8552719&288.971649&44.624531&12.160&5689.00&4.440&300.1&3&4.61$\pm$0.27&88.407030&&&&Keck&K&\\
K01800.01&11017901&285.268585&48.560009&12.394&5600.00&4.430&307.9&1&6.20$\pm$2.10&7.794300&&&&Keck&K&\checkmark\\
K01805.02&4644952&288.811951&39.770660&13.828&5708.00&4.080&789.5&3&4.30$\pm$1.20&31.782260&&&&Keck&K&\checkmark\\
K01805.01&4644952&288.811951&39.770660&13.828&5708.00&4.080&789.5&3&5.60$\pm$1.50&6.941344&&&&Keck&K&\checkmark\\
K01808.01&7761918&294.743317&43.461208&12.487&6278.00&4.350&424.4&1&4.00$\pm$1.60&89.192840&&&&Palomar&K&\checkmark\\
K01812.01&6279974&290.126526&41.601082&13.742&6285.00&4.420&778.7&1&4.80$\pm$1.80&0.805263&&&&Keck&K&\checkmark\\
K01825.01&5375194&295.300079&40.556591&13.895&5545.00&4.060&904.3&1&4.20$\pm$1.10&13.522604&&&&Palomar&K&\checkmark\\
K02672.01&11253827&296.132812&48.977402&11.921&5565.00&4.330&236.0&2&5.30$\pm$2.10&88.516580&&&&Palomar&K&\\
K02674.01&8022489&289.651245&43.824421&13.349&5973.00&4.260&501.1&3&7.30$\pm$2.60&197.510340&&&&Palomar&K&\checkmark\\
K02677.01&9958387&294.757690&46.831120&13.460&6409.00&4.340&694.7&1&6.50$\pm$1.00&237.788200&&&&Palomar&K&\checkmark\\
K03444.02&5384713&297.429199&40.561909&13.693&3842.00&4.664&122.4&4&5.74$\pm$3.20&60.326632&&&&KeckPalomar&JK&\checkmark\\
K03663.01&12735740&289.763611&51.962601&12.620&6007.00&4.340&394.8&1&11.29$\pm$0.03&282.525503&&&&&&\\
K03678.01&4150804&289.791595&39.285328&12.888&5650.00&4.313&413.8&1&9.12$\pm$0.04&160.885644&&&&Palomar&K&\checkmark\\
K03787.01&7813039&288.439117&43.505249&13.891&5993.00&4.355&741.5&1&9.17$\pm$1.10&141.733971&&&&Palomar&K&\checkmark\\
K03791.02&5437945&288.474854&40.651360&13.771&6340.00&4.163&932.2&2&5.65$\pm$0.21&220.130023&&&&Keck&JK&\\
K03791.01&5437945&288.474854&40.651360&13.771&6340.00&4.163&932.2&2&7.21$\pm$0.11&440.785197&&&&Keck&JK&\\
K03811.01&4638237&286.153107&39.714901&13.906&5551.00&4.518&505.5&1&6.98$\pm$0.99&290.140253&&&&Palomar&K&\checkmark\\
K03823.01&4820550&286.771454&39.983822&13.922&5817.00&4.544&663.9&1&5.54$\pm$0.08&202.117797&&&&Palomar&K&\checkmark\\
K03875.01&11911561&290.149139&50.239178&13.579&6022.00&4.027&660.4&1&4.73$\pm$0.70&8.870420&&&&Keck&K&\checkmark\\
K03907.01&7137213&296.997803&42.653320&12.642&6498.00&4.081&610.0&1&4.05$\pm$0.88&28.643425&&&&KeckPalomar&JHK&\checkmark\\
K05515.01&8429817&291.452271&44.431751&13.952&6247.00&4.255&806.9&1&8.88$\pm$1.50&6.263490&&&&Keck&JHK&\checkmark\\

\enddata

\end{deluxetable}

\begin{deluxetable}{lccccccccl}
\tabletypesize{\tiny}
\setlength{\tabcolsep}{0.03in}
\tablewidth{0pt}
\tablecaption{Visual companion detections with AO data.\label{tab:AO_params}}
\tablehead{
\colhead{\textbf{KOI}} &
\colhead{\textbf{$\Delta$ Mag}} &
\multicolumn{2}{c}{\textbf{Separation}} &
\multicolumn{2}{c}{\textbf{Distance}} &
\colhead{\textbf{Detection}} &
\colhead{\textbf{PA}} &
\colhead{\textbf{Association}} &
\colhead{\textbf{ref.}} \\
\colhead{\textbf{}} &
\colhead{\textbf{}} &
\colhead{\textbf{}} &
\colhead{\textbf{}} &
\colhead{\textbf{Primary}} &
\colhead{\textbf{Secondary}} &
\colhead{\textbf{Significance}} &
\colhead{\textbf{}} &
\colhead{\textbf{Probability}} &
\colhead{\textbf{}} \\
\colhead{\textbf{}} &
\colhead{\textbf{(mag)}} &
\colhead{\textbf{(arcsec)}} &
\colhead{\textbf{(AU)}} &
\colhead{\textbf{(pc)}} &
\colhead{\textbf{(pc)}} &
\colhead{\textbf{($\sigma$)}} &
\colhead{\textbf{(deg)}} &
\colhead{\textbf{}} &
\colhead{\textbf{}} \\
}

\startdata


\multirow{6}{*}{K00001} & 4.0 ($i$) & 1.13 & \nodata & \nodata &$259.0^{+1915.3}_{-231.2}$& \nodata & 135.0 &\nodata& L14 \\
 & 4.3 ($r$) & 1.11 & \nodata & \nodata &\nodata& \nodata & 135.5 &0.98& H14 \\
 & 3.3 ($z$) & 1.11 & \nodata & \nodata &\nodata& \nodata & 136.3 &0.99& H14 \\
 & 2.8 (J) & 1.12 & 232.6 & $207.0^{+22.4}_{-29.2}$ &\nodata& 233.5 & 136.2 &1.00& this work \\
 & 2.5 (H) & 1.11 & 230.5 & $207.0^{+22.4}_{-29.2}$ &\nodata& 383.6 & 136.3 &1.00& this work \\
 & 2.4 (K) & 1.12 & 230.9 & $207.0^{+22.4}_{-29.2}$ &\nodata& 68.5 & 136.5 &1.00& this work \\
\hline
K00005 & 2.3 (K) & 0.14 & 40.3 & $286.6^{+71.1}_{-15.8}$ &\nodata& 19.2 & 308.9 &1.00& CFOP \\
\hline
{\color{red}K00010$^\ast$} & 7.7 (J) & 3.04 & 2795.9 & $919.7^{+96.5}_{-126.8}$ &\nodata& \nodata & 94.3 &0.00& A12 \\
\hline
K00010 & 6.2 (J) & 3.74 & 3439.7 & $919.7^{+96.5}_{-126.8}$ &\nodata& 14.9 & 89.3 &0.15& A12 \\
\hline
K00017 & 3.8 (J) & 4.01 & 1982.1 & $494.3^{+42.6}_{-65.0}$ &\nodata& 132.6 & 39.9 &0.75& A12 \\
\hline
{\color{red}K00020$^\ast$} & 7.9 (J) & 5.04 & 3066.3 & $608.4^{+135.9}_{-20.1}$ &\nodata& \nodata & 139.6 &0.00& A12 \\
\hline
\multirow{2}{*}{K00097} & 4.0 (J) & 1.90 & 1500.4 & $789.7^{+56.9}_{-119.2}$ &$3847.0^{+2111.4}_{-1270.0}$& 49.8 & 105.1 &0.90& A12 \\
 & 4.0 (K) & 1.91 & 1508.3 & $789.7^{+56.9}_{-119.2}$ &\nodata& 81.0 & 105.1 &0.89& A12 \\
\hline
\multirow{5}{*}{K00098}  & 0.1 ($i$) & 0.29 & \nodata & \nodata &$707.4^{+1094.4}_{-533.0}$& \nodata & 140.0 &\nodata& L14 \\
& 0.5 ($r$) & 0.29 & \nodata & \nodata &\nodata& \nodata & 140.0 &1.00& H14 \\
& 0.7 ($z$) & 0.29 & \nodata & \nodata &\nodata& \nodata & 140.3 &1.00& H14 \\
 & 0.3 (J) & 0.27 & 120.5 & $446.2^{+109.2}_{-29.0}$ &\nodata& \nodata & 143.7 &1.00& A12 \\
 & 0.4 (K) & 0.28 & 124.9 & $446.2^{+109.2}_{-29.0}$ &\nodata& 15.7 & 143.5 &1.00& A12 \\
\hline
\multirow{2}{*}{K00098}  & 6.2 (J) & 5.60 & 2498.7 & $446.2^{+109.2}_{-29.0}$ &$2604.5^{+344.9}_{-930.6}$& 20.8 & 306.3 &0.13& A12\\
 & 5.6 (K) & 5.59 & 2494.3 & $446.2^{+109.2}_{-29.0}$ &\nodata& 12.0 & 306.1 &0.16& A12\\
\hline
\multirow{2}{*}{K00098}  & 7.2 (J) & 6.20 & 2766.1 & $446.2^{+109.2}_{-29.0}$ &$422.4^{+609.7}_{-139.4}$& 9.5 & 237.9 &0.00& this work\\
 & 6.3 (K) & 6.21 & 2769.5 & $446.2^{+109.2}_{-29.0}$ &\nodata& 5.1 & 237.9 &0.24& this work\\
\hline
{\color{red}K00108$^\ast$} & 7.2 (J) & 2.44 & 865.2 & $354.6^{+45.4}_{-39.2}$ &\nodata& \nodata & 74.9 &0.35& A12 \\
\hline
{\color{red}K00108$^\ast$} & 7.2 (J) & 4.87 & 1726.9 & $354.6^{+45.4}_{-39.2}$ &\nodata& \nodata & 112.4 &0.00& A12 \\
\hline
\multirow{3}{*}{K00119}  & 0.9 ($i$) & 1.05 & \nodata & \nodata &$216.1^{+302.3}_{-77.2}$& \nodata & 118.0 &\nodata& L14 \\
 & 0.2 (J) & 1.05 & 327.5 & $313.0^{+106.8}_{-62.2}$ &\nodata& 116.2 & 118.4 &0.60& this work \\
 & 0.2 (K) & 1.05 & 327.5 & $313.0^{+106.8}_{-62.2}$ &\nodata& 96.7 & 118.4 &1.00& this work\\
\hline
K00137 & 4.1 (J) & 5.59 & 2359.2 & $421.9^{+45.0}_{-54.0}$ &\nodata& 116.7 & 349.8 &0.57& A12\\
\hline
{\color{red}K00137$^\ast$} & 7.9 (J) & 4.80 & 2025.1 & $421.9^{+45.0}_{-54.0}$ &\nodata& \nodata & 340.5 &0.00& A12 \\
\hline
{\color{red}K00137$^\ast$} & 7.5 (J) & 4.98 & 2101.1 & $421.9^{+45.0}_{-54.0}$ &\nodata& \nodata & 136.3 &0.00& A12 \\
\hline
\multirow{3}{*}{K00141} & 1.4 ($i$) & 1.10 & \nodata & \nodata &$957.8^{+1256.9}_{-339.4}$& \nodata & 11.0 &\nodata& L14 \\
& 1.2 (J) & 1.06 & 490.9 & $463.1^{+32.7}_{-63.3}$ &\nodata& 192.1 & 13.9 &0.99& A12\\
 & 1.4 (K) & 1.06 & 490.9 & $463.1^{+32.7}_{-63.3}$ &\nodata& 242.4 & 13.5 &0.99& A12\\
\hline
K00152 & 5.7 (K) & 2.49 & 2230.1 & $897.1^{+162.0}_{-162.4}$ &\nodata& 6.0 & 29.5 &0.23& this work \\
\hline
K00157 & 4.4 (K) & 1.36 & 698.7 & $514.8^{+115.5}_{-93.5}$ &\nodata& 6.0 & 179.9 &0.94& this work \\
\hline
K00157 & 4.7 (K) & 4.09 & 2104.3 & $514.8^{+115.5}_{-93.5}$ &\nodata& 7.7 & 238.0 &0.35& this work \\
\hline
\multirow{3}{*}{K00279} & 3.5 ($r$) & 0.92 & \nodata & \nodata &$457.0^{+137.2}_{-53.4}$& \nodata & 246.6 &0.99& H14 \\
& 3.1 ($z$) & 0.92 & \nodata & \nodata &\nodata& \nodata & 247.2 &0.99& H14 \\
 & 2.3 (K) & 0.93 & 250.4 & $268.6^{+187.6}_{-46.3}$ &\nodata& 111.8 & 246.9 &1.00& CFOP\\
\hline
K00340 & 5.4 (K) & 5.34 & 2227.7 & $417.3^{+49.8}_{-46.5}$ &\nodata& 6.0 & 57.6 &0.10& this work \\
\hline
K00344 & 3.5 (K) & 4.13 & 2470.2 & $597.5^{+290.0}_{-126.2}$ &\nodata& 45.9 & 178.9 &0.68& this work  \\
\hline
K00344 & 5.2 (K) & 3.55 & 2123.6 & $597.5^{+290.0}_{-126.2}$ &\nodata& 9.7 & 210.6 &0.34& this work \\
\hline
K00366 & 6.5 (K) & 7.00 & 2464.4 & $352.1^{+468.6}_{-91.7}$ &\nodata& 5.6 & 70.1 &0.02& this work \\
\hline
{\color{red}K00372$^\ast$} & 8.6 (K) & 2.49 & 884.4 & $355.2^{+12.2}_{-54.0}$ &\nodata& \nodata & 157.8 &0.00& A12 \\
\hline
{\color{red}K00372$^\ast$} & 8.0 (K) & 3.56 & 1264.5 & $355.2^{+12.2}_{-54.0}$ &\nodata& \nodata & 56.9 &0.00& A12 \\
\hline
{\color{red}K00372$^\ast$} & 8.2 (K) & 4.99 & 1772.4 & $355.2^{+12.2}_{-54.0}$ &\nodata& \nodata & 170.7 &0.00& A12 \\
\hline
{\color{red}K00372$^\ast$} & 4.0 (K) & 5.94 & 2109.9 & $355.2^{+12.2}_{-54.0}$ &\nodata& \nodata & 32.7 &0.64& A12 \\
\hline
K00375 & 3.3 (K) & 5.47 & 4254.0 & $778.2^{+64.2}_{-139.5}$ &\nodata& 25.4 & 157.0 &0.64& this work \\
\hline
K00375 & 4.6 (K) & 3.19 & 2486.2 & $778.2^{+64.2}_{-139.5}$ &\nodata& 5.9 & 305.5 &0.56& this work \\
\hline
\multirow{2}{*}{K00377} & 4.5 (J) & 5.90 & 3654.5 & $619.4^{+54.5}_{-102.7}$ &$2639.5^{+578.1}_{-308.6}$& 37.5 & 91.7 &0.31& A12 \\
 & 4.2 (K) & 5.89 & 3648.3 & $619.4^{+54.5}_{-102.7}$ &\nodata& 101.0 & 91.7 &0.34& A12 \\
\hline
{\color{red}K00377$^\ast$} & 6.8 (J) & 2.79 & 1728.1 & $619.4^{+54.5}_{-102.7}$ &$9717.3^{+2629.4}_{-1745.8}$& \nodata & 37.9 &0.10& A12 \\
K00377 & 6.6 (K) & 2.79 & 1728.1 & $619.4^{+54.5}_{-102.7}$ &\nodata& 10.9 & 37.8 &0.02& A12 \\
\hline
K00633 & 3.9 (K) & 0.67 & 432.0 & $640.1^{+612.4}_{-71.0}$ &\nodata& 18.2 & 18.4 &0.97& this work \\
\hline
K00683 & 4.0 (K) & 3.42 & 2214.4 & $647.6^{+69.5}_{-92.2}$ &\nodata& 8.2 & 268.9 &0.32& this work \\

\enddata

\tablecomments{$\ast$: stellar companions that are not detected by our detection pipeline. References: A12 - \citet{Adams2012}; D14 - \citet{Dressing2014}; H14 - \citet{Horch2014}; L14 - \citet{Law2014}; LB14 - \citet{LilloBox2014}.}

\end{deluxetable}

\begin{deluxetable}{lccccccccl}
\tabletypesize{\tiny}
\setlength{\tabcolsep}{0.03in}
\tablewidth{0pt}
\tablecaption{Visual companion detections with AO data.\label{tab:AO_params}}
\tablehead{
\colhead{\textbf{KOI}} &
\colhead{\textbf{$\Delta$ Mag}} &
\multicolumn{2}{c}{\textbf{Separation}} &
\multicolumn{2}{c}{\textbf{Distance}} &
\colhead{\textbf{Detection}} &
\colhead{\textbf{PA}} &
\colhead{\textbf{Association}} &
\colhead{\textbf{ref.}} \\
\colhead{\textbf{}} &
\colhead{\textbf{}} &
\colhead{\textbf{}} &
\colhead{\textbf{}} &
\colhead{\textbf{Primary}} &
\colhead{\textbf{Secondary}} &
\colhead{\textbf{Significance}} &
\colhead{\textbf{}} &
\colhead{\textbf{Probability}} &
\colhead{\textbf{}} \\
\colhead{\textbf{}} &
\colhead{\textbf{(mag)}} &
\colhead{\textbf{(arcsec)}} &
\colhead{\textbf{(AU)}} &
\colhead{\textbf{(pc)}} &
\colhead{\textbf{(pc)}} &
\colhead{\textbf{($\sigma$)}} &
\colhead{\textbf{(deg)}} &
\colhead{\textbf{}} &
\colhead{\textbf{}} \\
}

\startdata


\multirow{3}{*}{K00697} & 0.0 (J) & 0.66 & 868.4 & $1315.7^{+209.8}_{-727.9}$ &$671.3^{+1881.0}_{-424.2}$& 235.5 & 55.3 &1.00& this work \\
 & 0.0 (H) & 0.66 & 868.4 & $1315.7^{+209.8}_{-727.9}$ &\nodata& 257.4 & 55.2 &1.00& this work \\
 & 0.0 (K) & 0.66 & 868.4 & $1315.7^{+209.8}_{-727.9}$ &\nodata& 257.1 & 55.7 &1.00& this work \\
\hline
\multirow{2}{*}{K01274} & 3.8 ($i$) & 1.10 & \nodata & \nodata &$412.2^{+81.1}_{-96.6}$& \nodata & 241.0 &\nodata& L14 \\
 & 2.4 (K) & 1.09 & 389.0 & $356.9^{+34.1}_{-55.8}$ &\nodata& 9.0 & 242.7 &0.99& this work \\
\hline
K01335 & 4.6 (K) & 4.51 & 6244.0 & $1385.0^{+209.1}_{-635.9}$ &\nodata& 17.3 & 358.1 &0.26& this work \\
\hline
K01353 & 5.4 (K) & 3.17 & 3466.8 & $1094.5^{+403.6}_{-410.2}$ &\nodata& 7.0 & 63.3 &0.19& this work \\
\hline
K01353 & 5.9 (K) & 5.65 & 6186.2 & $1094.5^{+403.6}_{-410.2}$ &\nodata& 5.0 & 97.9 &0.00& this work \\
\hline
\multirow{4}{*}{K01375} & 4.4 ($i$) & 0.77 & \nodata & \nodata &$1907.7^{+12041.7}_{-557.2}$& \nodata & 269.0 &\nodata& L14 \\
 & 3.8 (J) & 0.80 & 473.2 & $594.0^{+675.6}_{-94.2}$ &\nodata& 23.6 & 270.0 &0.97& this work \\
 & 3.6 (H) & 0.79 & 467.5 & $594.0^{+675.6}_{-94.2}$ &\nodata& 48.7 & 269.5 &0.98& this work \\
 & 3.6 (K) & 0.79 & 470.3 & $594.0^{+675.6}_{-94.2}$ &\nodata& 26.4 & 269.9 &0.98& this work \\
\hline
K01411 & 5.3 (K) & 3.79 & 1904.2 & $502.0^{+43.3}_{-62.1}$ &\nodata& 7.1 & 147.4 &0.14& this work \\
\hline
K01463 & 6.4 (K) & 6.11 & 2122.1 & $347.5^{+358.6}_{-73.1}$ &\nodata& 5.4 & 233.7 &0.03& this work \\
\hline
\multirow{3}{*}{K01784} & 0.9 (J) & 0.28 & 160.7 & $574.0^{+94.1}_{-117.7}$ &$575.8^{+2878.2}_{-445.1}$& 16.8 & 288.4 &1.00& this work \\
 & 0.8 (H) & 0.28 & 160.7 & $574.0^{+94.1}_{-117.7}$ &\nodata& 10.6 & 286.6 &1.00& this work \\
 & 0.7 (K) & 0.28 & 160.7 & $574.0^{+94.1}_{-117.7}$ &\nodata& 6.6 & 291.0 &1.00& this work \\
\hline
K01808 & 3.3 (K) & 4.68 & 1986.0 & $424.4^{+177.3}_{-70.8}$ &\nodata& 93.4 & 163.1 &0.74& this work \\
\hline
\multirow{2}{*}{K01812} & 4.3 ($i$) & 2.37 & \nodata & \nodata &$2173.3^{+254.0}_{-182.1}$& \nodata & \nodata &\nodata& LB14 \\
 & 3.6 (K) & 2.37 & 1844.2 & $778.7^{+168.5}_{-139.5}$ &\nodata& 21.6 & 297.9 &0.82& this work \\
\hline
K01825 & 4.4 (K) & 5.56 & 5025.4 & $904.3^{+133.3}_{-388.6}$ &\nodata& 25.9 & 295.3 &0.20& this work \\
\hline
K02672 & 3.5 (K) & 0.69 & 163.6 & $236.0^{+126.7}_{-46.5}$ &\nodata& 44.2 & 302.9 &0.99& CFOP \\
\hline
\multirow{2}{*}{K02672}  & 5.9 (K) & 4.54 & \nodata & \nodata &\nodata& \nodata & 308.0 &\nodata& D14 \\
 & 6.0 (K) & 4.61 & 1088.0 & $236.0^{+126.7}_{-46.5}$ &\nodata& 10.9 & 310.4 &0.23& this work \\
\hline
\multirow{4}{*}{K03444} & 2.8 ($i$) & 1.08 & \nodata & \nodata &$179.4^{+372.5}_{-128.1}$& \nodata & 9.6 &\nodata& LB14 \\
 & 2.6 ($z$) & 1.08 & \nodata & \nodata &\nodata& \nodata & 9.6 &0.99& LB14 \\
 & 2.2 (J) & 1.08 & 132.2 & $122.4^{+24.9}_{-27.1}$ &\nodata& 12.3 & 9.5 &1.00& this work \\
 & 2.4 (K) & 1.08 & 132.2 & $122.4^{+24.9}_{-27.1}$ &\nodata& 40.7 & 9.5 &1.00& this work \\
\hline
\multirow{4}{*}{K03444} & 4.5 ($i$) & 3.58 & \nodata & \nodata &$1878.9^{+3063.4}_{-812.6}$& \nodata & 264.4 &\nodata& LB14 \\
& 4.7 ($z$) & 3.58 & \nodata & \nodata &\nodata& \nodata & 264.4 &0.71& LB14 \\
 & 5.0 (J) & 3.63 & 443.8 & $122.4^{+24.9}_{-27.1}$ &\nodata& 11.5 & 264.8 &0.63& this work \\
 & 5.3 (K) & 3.57 & 436.7 & $122.4^{+24.9}_{-27.1}$ &\nodata& 26.9 & 264.8 &0.58& this work \\
\hline
K03678 & 3.3 (K) & 2.61 & 1081.7 & $413.8^{+200.9}_{-57.5}$ &\nodata& 33.9 & 169.6 &0.85& this work \\
\hline
K03787 & 5.1 (K) & 6.96 & 5162.8 & $741.5^{+328.6}_{-132.0}$ &\nodata& 9.2 & 254.9 &0.12& this work \\
\hline
K03823 & 5.6 (K) & 2.33 & 1546.0 & $663.9^{+233.3}_{-87.5}$ &\nodata& 10.0 & 58.0 &0.36& this work \\
\hline
K03823 & 5.1 (K) & 5.06 & 3357.5 & $663.9^{+233.3}_{-87.5}$ &\nodata& 14.9 & 239.4 &0.12& this work \\
\hline
\multirow{3}{*}{K03907} & 2.5 (J) & 2.77 & 1689.5 & $610.0^{+375.2}_{-135.7}$ &$588.4^{+3714.2}_{-521.8}$& 55.0 & 74.2 &0.95& this work \\
 & 2.2 (H) & 2.75 & 1674.6 & $610.0^{+375.2}_{-135.7}$ &\nodata& 140.3 & 74.3 &0.94& this work \\
 & 2.1 (K) & 2.76 & 1681.5 & $610.0^{+375.2}_{-135.7}$ &\nodata& 97.2 & 74.1 &0.96& this work \\
\hline
\multirow{3}{*}{K03907} & 4.3 (J) & 1.59 & 968.7 & $610.0^{+375.2}_{-135.7}$ &$99.5^{+1267.3}_{-29.0}$& 10.6 & 162.3 &0.93& this work \\
 & 3.7 (H) & 1.57 & 958.0 & $610.0^{+375.2}_{-135.7}$ &\nodata& 41.0 & 163.4 &0.95& this work \\
 & 3.5 (K) & 1.57 & 955.2 & $610.0^{+375.2}_{-135.7}$ &\nodata& 27.9 & 163.1 &0.96& this work \\
\hline
\multirow{3}{*}{K05515} & 4.1 (J) & 2.36 & 1907.5 & $806.9^{+288.9}_{-138.7}$ &$2059.3^{+13563.5}_{-1925.1}$& 24.3 & 297.9 &0.79& this work \\
 & 3.6 (H) & 2.36 & 1907.4 & $806.9^{+288.9}_{-138.7}$ &\nodata& 44.3 & 297.7 &0.82& this work \\
 & 3.7 (K) & 2.37 & 1915.9 & $806.9^{+288.9}_{-138.7}$ &\nodata& 23.2 & 297.7 &0.80& this work \\
\hline
\multirow{3}{*}{K05515}  & 5.4 (J) & 2.67 & 2158.2 & $806.9^{+288.9}_{-138.7}$ &$4443.6^{+25325.9}_{-4134.8}$& 9.1 & 114.2 &0.47& this work \\
 & 5.1 (H) & 2.68 & 2159.2 & $806.9^{+288.9}_{-138.7}$ &\nodata& 12.9 & 114.4 &0.45& this work \\
 & 5.1 (K) & 2.67 & 2155.6 & $806.9^{+288.9}_{-138.7}$ &\nodata& 6.4 & 114.5 &0.41& this work \\

\enddata

\tablecomments{$\ast$: stellar companions that are not detected by our detection pipeline. References: A12 - \citet{Adams2012}; D14 - \citet{Dressing2014}; H14 - \citet{Horch2014}; L14 - \citet{Law2014}; LB14 - \citet{LilloBox2014}.}

\end{deluxetable}

\end{document}